\documentclass{elsart}
\usepackage{graphicx}
\usepackage{amsmath}
\usepackage{amssymb}
\usepackage{amstext}
\begin{document}
\begin{frontmatter}
\title{Optimization of artificial flockings by means of anisotropy measurements}
\author{Motohiro Makiguchi and Jun-ichi Inoue}
\ead{makiguchi@complex.ist.hokudai.ac.jp, j$\underline{\,\,\,}$inoue@complex.ist.hokudai.ac.jp}
\address{
Complex Systems Engineering, 
Graduate School of Information Science and 
Technology,  
Hokkaido University, N14-W9, Kita-Ku, Sapporo 060-0814, Japan}
\begin{abstract}
An effective procedure to determine the optimal parameters appearing in 
artificial flockings is proposed in terms of optimization problems. 
We numerically examine genetic algorithms (GAs) to determine the optimal set of 
parameters such as the weights for three essential interactions in BOIDS  by Reynolds (1987)  
under `zero-collision'  and `no-breaking-up' constraints. 
As a fitness function (the energy function) to be maximized by the GA, we 
choose the so-called the $\gamma$-value of anisotropy which can be  
observed  empirically in typical flocks of starling. 
We confirm that the GA successfully finds the solution having 
a large $\gamma$-value leading-up to a strong anisotropy.  
The numerical experience shows that the procedure might enable us to 
make more realistic and efficient artificial flocking of starling even in our personal computers.
We also evaluate two distinct types of 
interactions in agents, namely, 
metric and topological definitions of interactions. 
We confirmed
 that the topological definition can explain 
the empirical evidence much better than 
the metric definition does. 
\end{abstract}

\end{frontmatter}
\section{Introduction}
\label{sec:Intro}
Collective behaviour of interacting agents such as flying birds, moving insects or 
swimming fishes 
shows highly non-trivial properties. 
We sometimes find a kind of `beauty' in the quite counter-intuitive and 
fascinating phenomena \cite{modelling,Inada,Okubo}. 
If one wishes to deal with 
these ingredients by mathematically rigorous 
approach, we sometimes regard each of them as a simple `particle' 
without size and any specific shape.  
As a typical example of such `massive' interacting particle systems,   
a critical phenomenon of order-disorder phase transitions with 
`spontaneous symmetry breaking' in spatial structures of the so-called 
ferromagnetic Ising system 
has attracted much attention of physicists.  
Up to now, a huge number of numerical and analytical studies in order to figure it out have been done 
by theoretical physicists and mathematicians \cite{Landau}.

On the other hand, for the mathematical modelling of many-particle systems 
having interacting {\it intelligent} agents (animals), we also use some 
probabilistic models. 
For instance, 
in the research field of physics, 
Vicsek {\it et.al.} \cite{Vicsek} 
proposed a flocking dynamics 
having a simple rule, 
namely, 
a given particle (agent) driven with 
a constant absolute velocity 
at each time step 
assumes the average direction of 
motion of the particles (agents) 
in its neighbourhood of 
radius $r$ with some random perturbation added. 

In engineering, a simplest and effective algorithm called 
BOIDS  \cite{Reynolds,BOIDS} has been widely used not only in the field of computer graphics 
but also in various other research fields including ethology, 
control theory and so on. The BOIDS simulates the collective behaviour of animal flocks 
by taking into account only a few simple rules for each 
interacting {\it intelligent} agent. 
 
Recently, quite a lot of useful flocking algorithms inspired by the BOIDS 
were proposed by a combination of a velocity cooperation   
with a local {\it potential-driven field} (for instance, see \cite{Tanner,Tanner2}). 
Among these studies, Olfati-Saber \cite{Olfati} provided a remarkable 
framework for designing of scalable flocking
algorithms. His framework has three essential factors in the algorithm.
The first one is the same 
three essential rules as those in the BOIDS we mentioned just above. 
The second factor is the ability of avoidance of unexpected obstacles 
appearing on the path of flock's movement. 
The third and the most remarkable one is the ability for causing the flock 
to track the path of a single {\it virtual leader} by introducing a navigational
feedback forth to each agent, 
namely,  all agents in the flock are moving according to the information about the 
virtual leader.  However, up to now, 
nobody knows whether such a virtual leader actually exists in real flocks or not.   

Hence, it is very hard task for us to 
evaluate these modelings and also very difficult to judge whether
it behaves like 
 {\it realistic} or not due to 
a lack of enough empirical findings to be compared
\cite{Aoki,Su,Herbert}.

As we know from the above issue (doubt) as an example, 
one of the serious problems in studies of 
any artificial flocking (algorithm) is apparently a lack of empirical data to check the validity. 
Actually, there are few studies to compare the results 
of the flocking simulations
with the empirical data. 
Therefore, the following essential queries still have been left unsolved; 
\begin{itemize}
\item
What is a criterion to determine to what extent 
the flocks seem to be {\it realistic}? 
\item
Is there any quantity (statistics) to measure the {\it quality} of 
the artificial flocks? 
\item 
Is it possible for us to construct the mathematically defined `optimal'  BOIDS in computers?  
If possible, how does one design 
the {\it optimal BOIDS} in terms of some maximization (or minimization)  
principle of appropriate fitness functions? 
\end{itemize}
From the view point of `engineering', the above first two queries are somewhat  
not essential because their main goal is to build-up 
a useful algorithm based on the collective behaviour of agents. 
However, from the natural science view points, 
the difference between empirical evidence and the result of the 
simulation is the most important issue and the consistency is a guide to judge the validity 
of the computer modelling and 
simulation. On the other hand, the third query is very important for engineering 
to solve important problems in the real world 
by using the knowledge of such outstanding abilities of these {\it intelligent} flockings. 

Recently, Ballerini {\it et. al.} \cite{Ballerini}
succeeded in obtaining the data for such collective  
animal behaviour, namely, empirical data of 
starling flocks containing up to a few thousands members. 
They also pointed out 
that the angular density of the nearest neighbours in the 
flocks is not uniform but 
apparently biased (it is weaken) along the direction of the flock's motion. 

With their empirical data by hand, in the previous paper \cite{Makiguchi}, we 
examined the possibility of the BOIDS simulations 
to reproduce this {\it anisotropy} and we also  
investigated numerically the condition on which the anisotropy emerges. 

However, 
in our previous studies,  we checked only the existence of the anisotropy and 
did not check extensively the strength of the anisotropy which 
is measured by the $\gamma$-value. 
To make the matter worse, 
due to some technical limitations of 
computer simulations, 
we could not evaluate the $n$-th order nearest neighbouring 
dependence of the $\gamma$-value precisely. 
Namely,  due to the following three bottlenecks, we could not 
check the validity of the BOIDS simulations by means of 
the anisotropy measurements: 
\begin{itemize}
 \item {\bf Bottleneck 1 :}
 It is very hard to check whether the result comes from 
 {\it the nature of mathematical modelling} or 
 from {\it the choice of parameters appearing in the model}. 
 \item {\bf Bottleneck 2 :} 
 It is very difficult for us to generate aggregations  
 with a high $\gamma$-value 
 which have been observed in a lot of empirical findings. 
 \item {\bf Bottleneck 3 :}
 It is very difficult to 
 evaluate the $\gamma$-value precisely 
 for the $n$-th order nearest neighbour 
 due to the so-called {\it border bias}. 
 \end{itemize}
\mbox{}
In this paper, 
in order to overcome the above 
{\bf Bottleneck 1} and {\bf Bottleneck 2}, we propose and examine 
a genetic algorithm (GA) to maximize the $\gamma$-value which 
is implicitly regarded as a fitness 
function of the weights of essential three interactions, 
namely, {\it Cohesion}, 
{\it Alignment} and {\it Separation}, appearing 
in the BOIDS algorithm. 
By finding 
the optimal weights for the BOIDS, 
we expect that the $\gamma$-value 
for the `optimal'  BOIDS 
is enhanced by the appropriate choice of 
the weights.  
For the {\bf Bottleneck 3}, 
we propose a border-bias free  
procedure to evaluate the $\gamma$-value 
in the computer simulations. 

This paper is organized as follows. 
In the next section 2, 
we explain the concept of 
anisotropy measurement.  
The anisotropy distribution map and 
the measurement through the $\gamma$-value 
are also explained. 
In the next section 3, 
we explain the essential three interactions in the BOIDS. 
In section 4 and 5, we 
provide the set-up of  scale-lengths and 
time-scale in our computer simulations. 
In section 6, 
we show the result without optimization as a preliminary.  
In the next section 7, we mention that the selecting the interactions in the BOIDS 
is formulated as an optimization problem 
to maximize 
the $\gamma$-value as the cost function. 
In section 8, we explain the 
genetic algorithm to maximize the cost function and 
why we use the algorithm. 
The results are shown in the next section 9. 
In these modelling, we assume that 
each agent interacts with the other mates within 
a fixed range of the visual field. 
In this sense, the model should be referred to as 
{\it metric model}. 
On the other hand, one can consider the 
{\it topological model} in which 
each agent interacts with a fixed number of 
the mates. 
In section 10, 
we apply our procedure to 
design the optimal BOIDS for the topological model and 
compare the result with that of the metric model. 
We find that the topological model 
can reproduce the empirical finding 
much better than the metric model does. 
In section 11, 
we discuss the results and the last section is summary.  
\section{Anisotropy in real and artificial flockings}
In this section, we 
explain the concept of 
anisotropy in flockings 
originally proposed by 
Ballerini {\it et. al.} \cite{Ballerini} 
to evaluate the empirical data 
of starling flockings. 
For the emergence of the anisotropy, we 
evaluate the strength of the anisotropy by 
the measurement, that is, the $\gamma$-value. 
\subsection{Emergence of anisotropy}
\begin{figure}[!t]
\begin{center}
\includegraphics[width=0.9\linewidth, bb=0 0 428 200]{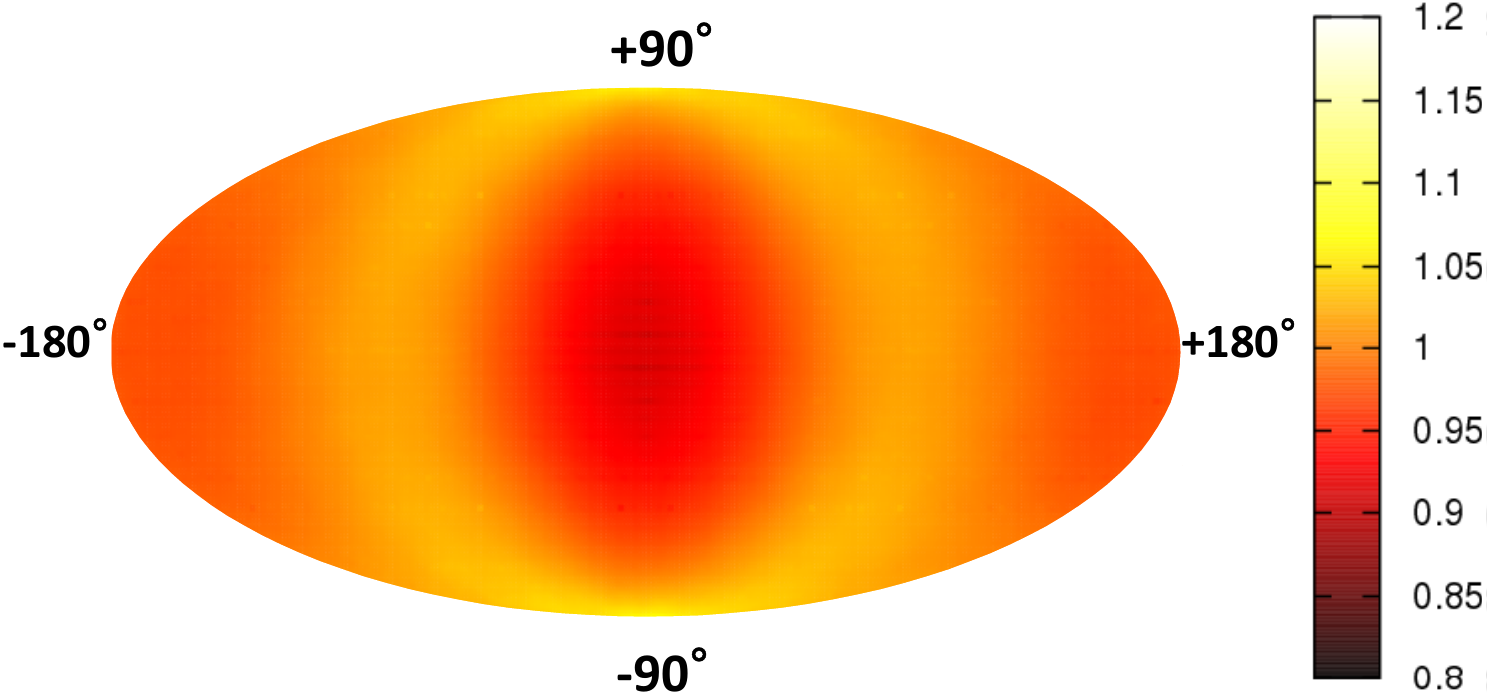}
\end{center}
 \caption{\footnotesize 
Angular distribution map simulated by BOIDS-based modelling  
(from Makiguchi and Inoue (2010) \cite{Makiguchi}).
}
\label{fig:fg1}
\end{figure}

Ballerini {\it et. al.} \cite{Ballerini} measured each bird's position 
 in the flocks  of starling (\textit{Sturnus vulgaris}) for $8$ seconds in three dimension.
To get such three dimensional data, they used `Stereo Matching'  which reconstructs 
three dimensional object
 from a set of stereo photographs. 
From these data, they calculated the angle between the direction of nearest 
neighbours and the direction
of the flock's motion for all birds in the flock. 
They measured the angles ($\phi$, $\alpha$), 
where $\phi$ 
 means the latitude ($\in [-90^\circ, 90^\circ]$) of nearest neighbour for each
bird measured from the direction of the flock's motion,
whereas the vertical axis $\alpha$ denotes longitude ($\in [-180^{\circ},180^{\circ}]$) 
which specifies the position of the nearest neighbour for each bird around the flock's motion, 
 of the nearest neighbour for all birds in the flock,
and plot these angles in the two-dimensional map using the so-called {\it Mollweide projection}.
To put it briefly, this map means the density of angular distribution of the nearest neighbour.
Inspired by their 
empirical findings, we 
simulate the distribution map by BOIDS simulation \cite{Makiguchi}. 
The resultant angular distribution map is shown in Fig. \ref{fig:fg1}.
This figure clearly shows that the density is not uniform but 
obviously biased.
\mbox{}\vspace{-0.4cm}
\subsection{Anisotropy measurement: Formula and generic properties}
\label{sec:gamma}
To evaluate the strength of the anisotropy, we 
use the $\gamma$-value which is an anisotropy measurement 
introduced by Ballerini {\it et. al.} \cite{Ballerini}. In following, we briefly explain 
how to compute it
and mention the general properties.
 
As a matter of convenience, let us first define the vector:  
\begin{equation}
\mbox{\boldmath $u$}_i^{(n)} \equiv  |u_{i}^{(n)} \rangle,\,\,\,
((\mbox{\boldmath $u$}_i^{(n)})^{t} =(u_{ix}^{(n)},
u_{iy}^{(n)},u_{iz}^{(n)}) \equiv  \langle u_{i}^{(n)}|)
\end{equation}
in the {\it Dirac's bracket representation} 
in quantum mechanics \cite{Dirac} as a three-dimensional unit vector pointing to the $n$-th nearest neighbouring 
agent (bird) from an arbitrary argent $i$.  

We should keep in mind that 
$t$ appearing in the shoulder of matrix here such as 
$\mbox{\boldmath $A$}^{t}$ stands for the `transpose'. 
Then, we  have the following 
$3 \times 3$ projection matrix 
$\mbox{\boldmath $M$}^{(n)}$  in terms of the Dirac's bracket:  
\begin{eqnarray}
\mbox{\boldmath $M$}^{(n)} & = & 
\frac{1}{N} \sum_{i=1}^{N}
(|u_{i}^{(n)} \rangle \langle u_{i}^{(n)}|)
\end{eqnarray}
whose components are given by 
\begin{eqnarray}
(\mbox{\boldmath $M$}^{(n)})_{\alpha \beta} & = & \frac{1}{N}\sum_{i=1}^{N}
(|u_{i}^{(n)}\rangle)_\alpha
(|u_{i}^{(n)}\rangle)_\beta 
\label{eq:M}
\end{eqnarray}
for  $\alpha, \beta=x, y, z$, 
where $N$ stands for the total number of 
agents in the flock. 
For the above $\mbox{\boldmath $M$}^{(n)}$, 
we immediately obtain 
the normalized eigenvectors $|U_{k}^{(n)} \rangle, k=1,2,3$,
and one can rewrite the matrix $\mbox{\boldmath $M$}^{(n)}$ in terms of 
these bases as 
\begin{eqnarray}
\mbox{\boldmath $M$}^{(n)} & = & 
\sum_{k=1}^{3}
\lambda_{k} |U_{k}^{(n)} \rangle \langle U_{k}^{(n)}| 
\end{eqnarray}
where $\lambda_{k}, k=1,2,3$ stand for 
the eigenvalues of the matrix, that is, 
 $\mbox{\boldmath $M$}^{(n)}|U_{k}^{(n)} \rangle = 
\lambda_{k} |U_{k}^{(n)} \rangle, k=1,2,3$, and 
of course, the rectangular condition $\langle U_{k}^{(n)} | U_{l}^{(l)} \rangle = 
\delta_{k,l}$ is satisfied. 
It should be noted that 
when the projection matrix $\mbox{\boldmath $M$}^{(n)}$ 
becomes irregular, we cancel the observation in our calculations of the $\gamma$-value. 
Therefore, 
the probability $P_{k}$ that 
an arbitrary agent exists in the direction of 
the vector $|U_{k}^{(n)} \rangle$ 
is explicitly given by 
\begin{eqnarray}
P_{k} & = & 
| \langle U_{k}^{(n)} |
\mbox{\boldmath $M$}^{(n)}|
U_{k}^{(n)} \rangle |^{2}=\lambda_{k}^{2}, \,\,\,
k=1,2,3. 
\end{eqnarray}
When we put these eigenvalues in 
a particular order, say,  $\lambda_{1} < \lambda_{2} < \lambda_{3}$ 
(this reads $P_{1}<P_{2}<P_{3}$), 
the vector $|U_{3}^{(n)} \rangle$ 
is the direction in which the agents are more likely to exist, 
whereas there are fewest agents in the direction of vector 
$|U_{1}^{(n)} \rangle$. 
Therefore, 
if we define the emergence of {\it anisotropy} as {\it the absence of the birds along the direction of the flock' s motion}, 
the strength of the anisotropy is naturally 
measured by the inner-product of 
the vector pointing to the flock's movement 
and the eigenvector having the lowest eigenvalue. 

To evaluate the anisotropy measurement more 
explicitly,  let us define the eigenvector 
of the lowest eigenstate by 
$|W^{(n)} \rangle$, 
that is, 
$\langle W^{(n)}| \mbox{\boldmath $M$}^{(n)}|W^{(n)} \rangle = 
\min_{k} \lambda_{k}$. 
Then, the anisotropic measurement is calculated from 
the $|W^{(n)} \rangle$ and 
the vector $|V \rangle $ that points to 
the direction of 
the flock's movement (the velocity of the center of mass 
in the flocking 
$|V \rangle \equiv 
(1/N)\sum_{i=1}^{N} 
\mbox{\boldmath $V$}_{i}$) 
as 
\begin{eqnarray}
\gamma_{t} & = & 
|\langle W^{(n)} | V \rangle|^{2}. 
\label{eq:defgamma}
\end{eqnarray}
\mbox{}

Obviously, the above $\gamma_{t}$ is 
dependent on the time $t$ through 
the time-dependence of $|V\rangle$ and 
$|W^{(n)}\rangle$. 
Thus, we define the anisotropic measurement $\gamma$ by 
averaging over the `observation-time' with the infinite length $T \to \infty$ as 
 \begin{eqnarray}
 \gamma & = & 
 \mathbb{E}_{t} [\gamma_{t}] \equiv 
 \lim_{T \to \infty}
 \frac{1}{T}\sum_{t=1}^{T}\gamma_{t}
 \label{eq:timeAve}.
 \end{eqnarray}
As it is impossible to take the infinite observation-time 
limit $T \to \infty$ in  computer, we replace the limit by finite observation-time, 
say, $T=80$ for the upper bound in the sum (\ref{eq:timeAve}). 

We should also keep in mind that 
the $\gamma$-value depends on the choice of 
initial conditions and one should take the average over 
the distribution of initial conditions. 
However, we might expect that 
the $\gamma$-value calculated for a single realization of 
initial conditions, say, $\gamma$ is identical to its average $\mathbb{E}_{ini.} [\cdots]$ over the initial condition, 
namely, $\gamma=\mathbb{E}_{ini.} [\gamma]$ in 
the limit of $N \to \infty$. 
As we shall show later, 
the number of agents in our simulation is too small 
$N=100 \ll \infty$ to satisfy the above condition. 
Therefore, we calculate the average of the $\gamma$-value  
for $1000$-independent initial conditions. 

It should be noted that the $\gamma$-value takes any positive values  
in the range $0 \leq \gamma \leq 1$.  
For the case of $\gamma=0$, 
the nearest neighbour 
is more likely to exist in 
the direction of flock's movement, 
namely, $\langle W^{(n)} | V \rangle=0$.  
On the other hand,  the $\gamma=1$ 
implies that 
the nearest neighbour 
exist in the two-dimensional 
plane which is perpendicular to 
the flock's movement with probability $1$, 
that is to say, $|W^{(n)}\rangle = \pm |V \rangle$ 
resulting in $|\langle W^{(n)}|V \rangle|^{2}=1$. 
We should notice that for the eigenvector $|S^{(n)} \rangle$ and  $|S_{2}^{(n)} \rangle$ 
having the largest and the second largest eigenvalues, 
$|\langle S^{(n)}|V \rangle|^{2}+|\langle S_{2}^{(n)}|V \rangle|^{2}=1$ for $\gamma=0$ and 
$\langle S^{(n)}|V \rangle=\langle S_{2}^{(n)}|V \rangle=0$ for $\gamma=1$ 
should be satisfied.   

After simple algebra, 
one can show that the $\gamma$ takes 
$1/3$ when there is no spatial bias 
in the direction of the nearest neighbours
(namely {\it isotropy}). %
This means that the anisotropy emerges 
when the following condition is satisfied.  
\begin{eqnarray}
\gamma & > & \gamma_{\rm uniform} \equiv \frac{1}{3} 
\label{eq:cond_gamma}
\end{eqnarray}
In fact, the above inequality is easily confirmed. 
The $\gamma$-value 
for the uniform distribution 
of the position $(\phi,\alpha)$, 
where $\phi$ and $\alpha$ are the same
 variables of %
angles defined in the previous subsection, 
for a given vector $|V \rangle$, 
namely, the $\gamma$-value for $\rho (\phi, \alpha) =(4\pi)^{-1}$ is 
easily calculated as 
\begin{eqnarray}
\gamma_{\rm uniform} & = &  
\int_{\rm sphere} 
\rho (\phi, \alpha)\, d\phi \,d\alpha 
|\langle W^{(n)}|V\rangle|^2 = \frac{1}{4 \pi}
\int_{-\pi}^{\pi}\cos^{2} \alpha \, d \alpha
\int_{-\frac{\pi}{2}}^{\frac{\pi}{2}}
\cos^{3}\phi \, d\phi \nonumber \\
\mbox{} & = &  \frac{1}{3}
\end{eqnarray}
where we used $|\langle W^{(n)}|V\rangle|^2 = \cos^{2} \phi \cos^{2} \alpha $. 
Therefore, the distribution of the $n$-th nearest neighbours has 
an anisotropic structure when the $\gamma$-value is larger than 
$\gamma_{\rm uniform} = 1/3$,
 namely the condition for the emergence of the anisotropy is 
 explicitly written by (\ref{eq:cond_gamma}).

Ballerini {\it et. al.} \cite{Ballerini} measured 
the $\gamma$-value up to the $n$-th order 
of nearest neighbours for 
two kinds of empirical data having different numbers of agents 
and show the $n$-dependence of 
the $\gamma$-value. 
From their plot, we clearly find that 
$\gamma$-value takes larger than $0.8$ 
for $n=1$ and the value remains larger than $\gamma_{\rm uniform} = 1/3$ up to $n=6$. 
In our previous studies \cite{Makiguchi}, 
the $n$-dependence of the 
$\gamma$-value was evaluated for the data generated artificially from 
BOIDS simulations, however, 
we could not overcome the problem of 
{\it border bias} which was mentioned as 
{\bf Bottleneck 3} in the previous section. 
In this paper, we introduce a way to 
overcome this technical difficulty and 
attempt to measure the $\gamma$-value 
more precisely. 
\section{Essential three interactions in BOIDS}
\label{sec:Model}
To make flock simulations in computer,
 we use the so-called BOIDS which was originally designed by Reynolds in 1987 \cite{Reynolds}.
The BOIDS is one of the well-known mathematical (probabilistic) models in the research fields of CG and animation. 
Actually, the BOIDS can simulate very complicated animal flocks or schools
although it consists of just only three simple interactions for each agent in the aggregation: 
\begin{enumerate}
 \item[(c)]{\bf Cohesion}: Making each agent's position 
 $\mbox{\boldmath $X$}_{i}\,(i=1,\cdots,N)$ toward the average position of neighbouring flock mates.
 \item[(a)]{\bf Alignment}: Keeping the velocity of each agent 
 $\mbox{\boldmath $V$}_{i}\,(i=1,\cdots,N)$ 
 the average value of neighbouring flock mates.
 \item[(s)]{\bf Separation}: Making a vector of each agent's position 
 $\mbox{\boldmath $X$}_{i}\,(i=1,\cdots,N)$ to avoid the collision with the neighbouring flock mates.
\end{enumerate}
Each agent decides her (or his) next direction of migration by compounding these three vectors of interaction.
In addition to this, 
it is important for us to bear in mind that 
`local flock mates' mentioned above denotes the neighbours within the range of view for each agent.
We explain this view and other settings of our simulation in the next section.
\subsection{BOIDS dynamics}
For simplicity, we define 
`neighbouring mates'  by 
all metes which exist within 
the visual field with a radius $R$ and 
for each mate categorized as the neighbouring mates, 
we calculate the interactions of 
{\it Cohesion} and {\it Alignment} as normalized unit vectors. 
In addition, we evaluate the interaction of 
{\it Separation} as a unit vector pointing to the direction 
of fading-out from the mates. 
Each agent $i$ updates its own velocity vector 
$\mbox{\boldmath $V$}_{i}$ and the position $\mbox{\boldmath $X$}_{i}$ by 
the following recursion relations. 
\begin{eqnarray}
\mbox{\boldmath $V$}_{i}(l+1) & = & 
\overline{V}_{l}^{(i)}
\mbox{\boldmath $e$}_{B}^{(i)}(l) 
\label{eq:BOIDS_metric} \\
\mbox{\boldmath $X$}_{i}(l+1) & = & 
\mbox{\boldmath $X$}_{i}(l) + \mbox{\boldmath $V$}_{i}(l+1)
\label{eq:BOIDS_metricX}
\end{eqnarray}
where 
$l$ denotes time step in our simulations 
and we discretized 
the infinitesimal 
time as a unit time step $\Delta l=l+1-l=1$ in 
the definition of velocity 
$\mbox{\boldmath $V$}_{i}= 
d\mbox{\boldmath $X$}_{i}/dl \simeq 
\{\mbox{\boldmath $X$}_{i}(l+\Delta l)-
\mbox{\boldmath $X$}_{i}(l)\}/\Delta l = 
 \mbox{\boldmath $X$}_{i}(l+1)-
 \mbox{\boldmath $X$}_{i}(l) \equiv \mbox{\boldmath $V$}_{i}(l+1)$ 
 to obtain (\ref{eq:BOIDS_metricX}). 
$\mbox{\boldmath $e$}_{B}^{(i)}(l)$ denotes a 
unit vector pointing to 
the direction to which the agent $i$ should move 
according to the BOIDS. 
The $\mbox{\boldmath $e$}_{B}^{(i)}(l)$ is explicitly given by 
\begin{eqnarray}
\mbox{\boldmath $e$}_{B}^{(i)}(l) & =& 
\frac{
\frac{J_{1}\mbox{\boldmath $v$}_{\rm C}^{(i)} (l)+ 
J_{2} \mbox{\boldmath $v$}_{\rm A}^{(i)}(l)
+J_{3}\mbox{\boldmath $v$}_{\rm S}^{(i)}(l)}
{|
J_{1}\mbox{\boldmath $v$}_{\rm C}^{(i)} (l)+ 
J_{2} \mbox{\boldmath $v$}_{\rm A}^{(i)}(l)
+J_{3}\mbox{\boldmath $v$}_{\rm S}^{(i)}(l)|} 
+ 
\eta 
 \frac{\mbox{\boldmath $V$}_{i}(l)}{|\mbox{\boldmath $V$}_{i}(l)|}}
{
\left|
\frac{J_{1}\mbox{\boldmath $v$}_{\rm C}^{(i)} (l)+ 
J_{2} \mbox{\boldmath $v$}_{\rm A}^{(i)}(l)
+J_{3}\mbox{\boldmath $v$}_{\rm S}^{(i)}(l)}
{|
J_{1}\mbox{\boldmath $v$}_{\rm C}^{(i)} (l)+ 
J_{2} \mbox{\boldmath $v$}_{\rm A}^{(i)}(l)
+J_{3}\mbox{\boldmath $v$}_{\rm S}^{(i)}(l)|} 
+ 
\eta 
\frac{\mbox{\boldmath $V$}_{i}(l)}{|\mbox{\boldmath $V$}_{i}(l)|}
\right|} 
\label{eq:BOIDS_metric2}
\end{eqnarray}
with 
\begin{eqnarray}
\mbox{\boldmath $v$}_{\rm C}^{(i)} (l) & = & 
\frac{
\frac{
\sum_{j=1}^{N} 
\Theta (R-r_{ij}) \mbox{\boldmath $X$}_{j}(l)
}{\sum_{j=1}^{N}\Theta (R-r_{ij})}
 -\mbox{\boldmath $X$}_{i}(l)}
{
\left|
\frac{\sum_{j=1}^{N}
\Theta (R-r_{ij}) 
\mbox{\boldmath $X$}_{j}(l)
}{
\sum_{j=1}^{N}\Theta (R-r_{ij})}
-\mbox{\boldmath $X$}_{i}(l)
\right|}  
\label{eq:VC_metric} \\
\mbox{\boldmath $v$}_{\rm A}^{(i)}(l) & = & 
\frac{
\sum_{j=1}^{N}
\Theta (R-r_{ij}) 
\mbox{\boldmath $V$}_{j}(l)}
{|
\sum_{j =1}^{N}
\Theta (R-r_{ij})  
\mbox{\boldmath $V$}_{j}(l)|} 
\label{eq:VA_metric} \\
\mbox{\boldmath $v$}_{\rm S}^{(i)} (l)  & = & 
-\frac{
\sum_{j =1}^{N}
\Theta (R-r_{ij})
(\mbox{\boldmath $X$}_{j}(l)-
\mbox{\boldmath $X$}_{i}(l))
}
{|
\sum_{j =1}^{N} 
\Theta (R-r_{ij})
(\mbox{\boldmath $X$}_{j}(l)-
\mbox{\boldmath $X$}_{i}(l))
|}
\label{eq:VS_metric}
\end{eqnarray}
where we defined 
$r_{ij}$ as the 
square distance between 
agent
$i$ and $j$ as 
\begin{eqnarray}
r_{ij} & \equiv &  
|\mbox{\boldmath $X$}_{i}(l)-
\mbox{\boldmath $X$}_{j}(l)|=
\sqrt{\{\mbox{\boldmath $X$}_{i}(l)-
\mbox{\boldmath $X$}_{j}(l)\}^{2}}.
\end{eqnarray}
$\Theta (\cdots)$ denotes a step function. 
Therefore, 
$\sum_{j=1}^{N}\Theta (R-r_{ij})$ stands for the 
number of `neighbouring mates' 
for the agent $i$ and 
the number is obviously dependent on the 
agent $i$. 

A balance parameter $\eta$ appearing in 
(\ref{eq:BOIDS_metric2}) 
determines 
the weights 
of  two distinct modifications 
$\mbox{\boldmath $e$}_{B}^{(i)}(l)$ 
for the velocity 
vector for the agent $i$, 
namely, 
the `BOIDS-driven' correction term 
$\sim\, J_{1}\mbox{\boldmath $v$}_{\rm C}^{(i)} (l)+ 
J_{2} \mbox{\boldmath $v$}_{\rm A}^{(i)}(l)
+J_{3}\mbox{\boldmath $v$}_{\rm S}^{(i)}(l)$
and the vector conservation term 
$\sim \, \mbox{\boldmath $V$}_{i}(l)$. 
Hence, the vector $\mbox{\boldmath $V$}_{i}(l)$ 
is conserved for $\eta \gg 1$, whereas 
the dynamics of 
$\mbox{\boldmath $V$}_{i}(l)$ becomes purely BOIDS-driven for $\eta \ll 1$. 
Therefore, the choice of $\eta$ is regarded as 
a kind of `inert effect' in the dynamics of BOIDS. 
The value itself should be determined empirically. 
However, due to the lack of such 
useful information about the inertia 
in real flockings, here we simply set $\eta=2$ by ad-hoc manner.

From the above definition of (\ref{eq:VC_metric}), 
we easily find that 
$\mbox{\boldmath $v$}_{\rm C}^{(i)}(l)=-
\mbox{\boldmath $v$}_{\rm S}^{(i)}(l)$ and 
one of these two distinct effects is 
completely cancelled in the BOIDS 
dynamics (\ref{eq:BOIDS_metric})(\ref{eq:BOIDS_metricX}) 
as $ \sim (J_{1}-J_{3})\mbox{\boldmath $v$}_{\rm C}^{(i)}(l)$ 
for any choice of $J_{1}, J_{3}$.
To correct  this undesirable situation, 
we slightly modify the 
$\mbox{\boldmath $v$}_{\rm S}^{(i)}(l)$ 
as follows. 
\begin{eqnarray}
\mbox{\boldmath $v$}_{\rm S}^{(i)} (l)  & = & 
-\frac{
\Theta (R_{0}-r_{i \overline{j(1:i)}})
(\mbox{\boldmath $X$}_{\overline{j(1: i)}}(l)-
\mbox{\boldmath $X$}_{i}(l))
}
{|
\Theta (R_{0}-r_{i \overline{j(1:i)}})
(\mbox{\boldmath $X$}_{\overline{j(1: i)}}(l)-
\mbox{\boldmath $X$}_{i}(l))
|}
\label{eq:VS_metric_mod}
\end{eqnarray}
where 
$\overline{j(n:i)}$ 
denotes the $n$-th nearest neighbouring mate 
of the agent $i$ and it is 
explicitly given by :
\begin{eqnarray}
\overline{j(n:i)} & \equiv & 
 {\rm argmax}_{j \neq \overline{j(n-1:i)},\cdots,\overline{j(0:i)}}
\,\,r_{ij}
\label{eq:def_nth_nn}
\end{eqnarray}
with $\overline{j (0:i)} \equiv i$. 
Then, $\overline{j(1:i)}$ means the nearest neighbouring site 
\begin{eqnarray}
\overline{j(1:i)} & \equiv & 
{\rm arg} \min_{j}r_{ij}
\end{eqnarray}
for each $i$. Namely, 
the separation $\mbox{\boldmath $v$}_{\rm S}^{(i)}(l)$ 
acts if and only if the distance 
between the agent $i$ and 
the nearest neighbouring 
mate $\overline{j(1:i)}$ is lower than the radius of 
separation range $R_{0}$. 

The vectors $ \mbox{\boldmath $v$}_{\rm C}^{(i)}(l),
\mbox{\boldmath $v$}_{\rm A}^{(i)}(l)$ 
and $\mbox{\boldmath $v$}_{\rm S}^{(i)}(l)$
 denote the components caused by 
 the interactions {\it Cohesion}, {\it Alignment} and 
{\it Separation} for agent $i$ at time $l$, respectively. 
We should keep in mind that 
$|\mbox{\boldmath $v$}_{\rm C}^{(i)}(l)|=
|\mbox{\boldmath $v$}_{\rm A}^{(i)}(l)|=
|\mbox{\boldmath $v$}_{\rm S}^{(i)}(l)|=1$ holds and 
$\mbox{\boldmath $J$}=(J_{1},J_{2},J_{3})$ 
stands for the set of weights for three interactions, 
namely, 
{\it Cohesion}, {\it Alignment} and 
{\it Separation}. 
We should keep in mind 
that 
from (\ref{eq:VC_metric}),
(\ref{eq:VA_metric}) 
and (\ref{eq:VS_metric}), 
the weight $J_{2}$ is a `dimension-less' variable, 
however, $J_{1}$ and $J_{3}$ have 
inverse-time dimension $\sim (\mbox{time step})^{-1}$. 

We should keep in mind that 
the above definition of 
$\overline{j(n:i)}$, the $\gamma$-value is calculated 
by setting 
$| u_{i}^{(n)} \rangle \equiv \mbox{\boldmath $u$}_{i}^{(n)}=
(\mbox{\boldmath $V$}_{i \overline{j(n:i)}}-\mbox{\boldmath $V$}_{i})/
{|\mbox{\boldmath $V$}_{i \overline{j(n:i)}}-\mbox{\boldmath $V$}_{i}|}$ 
in (\ref{eq:M})(\ref{eq:defgamma}) and (\ref{eq:timeAve}).

From equation (\ref{eq:BOIDS_metric}), 
we are confirmed that the amplitude of 
velocity vector of agent $i$ at time $l+1$ is identical to 
the average amplitude of 
velocity vectors for neighbouring mates in the 
previous time step $l$ as 
\begin{eqnarray}
|\mbox{\boldmath $V$}_{i}(l+1)| & = & 
\overline{V}_{l}^{(i)} \equiv 
\frac{
\sum_{j=1}^{N}
\Theta (R-r_{ij})
|\mbox{\boldmath $V$}_{j}(l)|}
{
\sum_{j=1}^{N}
\Theta (R-r_{ij})}.
\end{eqnarray}
The above update rules 
(\ref{eq:BOIDS_metric}),
(\ref{eq:BOIDS_metricX}),
(\ref{eq:BOIDS_metric2}), 
(\ref{eq:VC_metric}),
(\ref{eq:VA_metric}) and 
(\ref{eq:VS_metric_mod}) are our basic 
dynamical equations to be evaluated numerically. 

Obviously, the behaviour of 
the artificial flockings strongly depends on 
the choice of the weights, 
however, there is no extensive study 
to investigate to what extent the behaviour 
changes quantitatively by changing the weights. 
From the fact in mind, 
in this paper, we propose 
a systematic algorithm to 
determine 
the weight by using the evolutionary 
computation such as GAs to 
maximize the $\gamma$-value as a fitness function.  
\section{Scale-lengths in BOIDS simulations}
We here explain how we set several 
scale-lengths appearing in our simulations. 
In our previous studies, we determined them 
without any justification 
from the empirical evidence, however, 
here we attempt choose the 
scale lengths by taking into account the data 
from the reference \cite{Ballerini2} in order to 
realize the artificial flockings 
as {\it realistic} as possible
(We summarize these variables in Table \ref{tab:tb1}). 

However, some parameters is not determined by empirical data,
for instance, the range of interaction, 
the {\it Frame-Rate (FR)} [fps:frame-per-second] and so on.
Therefor we set these parameter by the subjective view point and 
some regards for the calculation cost, for example,
 the number of agents $N=100$, 
the radius of the visual field $R= 3 \times R_{0}$ 
where $R_{0}=1.09$ [m] denotes the radius of separation range and 
the $FR$, 
which will be explained in the next section in detail, 
is $200$ [Hz]. 

\begin{table}[!t]
\begin{center}
\begin{tabular}{l|c}
	\hline
	\hline
	\multicolumn{2}{c}{A set of scale-lengths in our BOIDS}\\
	\hline
	Number of agents ($N$) & 100\\
	Body-Length ($BL$) & 0.2 [m]\\
	Wing-Span ($WS$) & 0.4 [m] \\
	Radius of Separation Range ($R_0$) & 1.09 [m]\\
	Radius of Visual Field ($R$) & 3 $\times$ $R_0$ [m]\\
	Initial Speed Average ($V^{'}$) & 10.10 [m/s]\\
	Initial Density of the Aggregation ($\rho$) & 0.13 [${\rm m}^{-3}$] \\
	\hline
	\hline
\end{tabular}
\end{center}
\caption{\footnotesize 
A set of scale-lengths in our flock simulation.
Variables other than the {\it Number of agents} and
 the {\it Radius of Visual Field} are based on
empirical data by Ballerini {\it et. al.} 
({\sf Event 29-03} in Table 1 of \cite{Ballerini2}). 
}
\label{tab:tb1}
\end{table}
\section{On the time-scale in BOIDS simulations}
In our previous study \cite{Makiguchi}, 
we defined the unit time (frame) by $0.1$ [sec].
In this paper, we shall define the frame based 
on the so-called  Frame-Rate($FR=200$[Hz]). 
In order to consider the consistency with the 
empirical data analysis by Ballerini {\it et. al.} \cite{Ballerini} 
in which they used $0.1$ [sec] for a unit frame, 
we evaluate the $\gamma$-value every 
$FR/20$ frames and 
the distance covered by each agent per frame, 
that is, the average of flock's velocity $V$ is 
also determined from the empirical evidence 
of velocity $V^{'}$ [m/sec] as $V=V^{'}/FR$ [${\rm frame}^{-1}$]. 

In our previous work \cite{Makiguchi}, 
we also chose the initial velocity for 
each agent from a uniform distribution having a finite support. 
However, 
this procedure might cause 
some difficulties, 
namely, 
we might encounter the `breaking-up' of 
the flocking to several small groups due to synchronization in their speeds of convergence. 
In general, it is very difficult for us 
to control the speed of the flocking 
(the speed of the center of mass) 
after each agent's speed converges when 
we determine the initial speed of each agent by a random number 
from a uniform distribution. 
To overcome this type of difficulties, we 
sample the initial value of each agent's velocity $V_{i}$ from  
the following Gaussian with mean $V^{'}=10.10$ [m/sec]  and 
variance $\sigma^{2}=(V^{'}-1)/3$, namely, 
\begin{eqnarray}
P(V_{i}) & = & 
\frac{1}{\sqrt{2\pi} \sigma}
\,
{\exp}
\left[
-\frac{(V_{i}-V^{'})^{2}}{2\sigma^{2}}
\right].
\end{eqnarray}
By this setting, we are confirmed 
that the speed of flocking 
actually converges to $V^{'}$. 
\section{A preliminary: Simulations without GA}
In this section, we show the results 
without any searching of the optimal 
weights of interactions by genetic algorithms 
as a preliminary. 
\subsection{Preliminary results}
\begin{figure}[!t]
\begin{center}
\includegraphics[width=0.65\linewidth,bb=0 0 445 224]{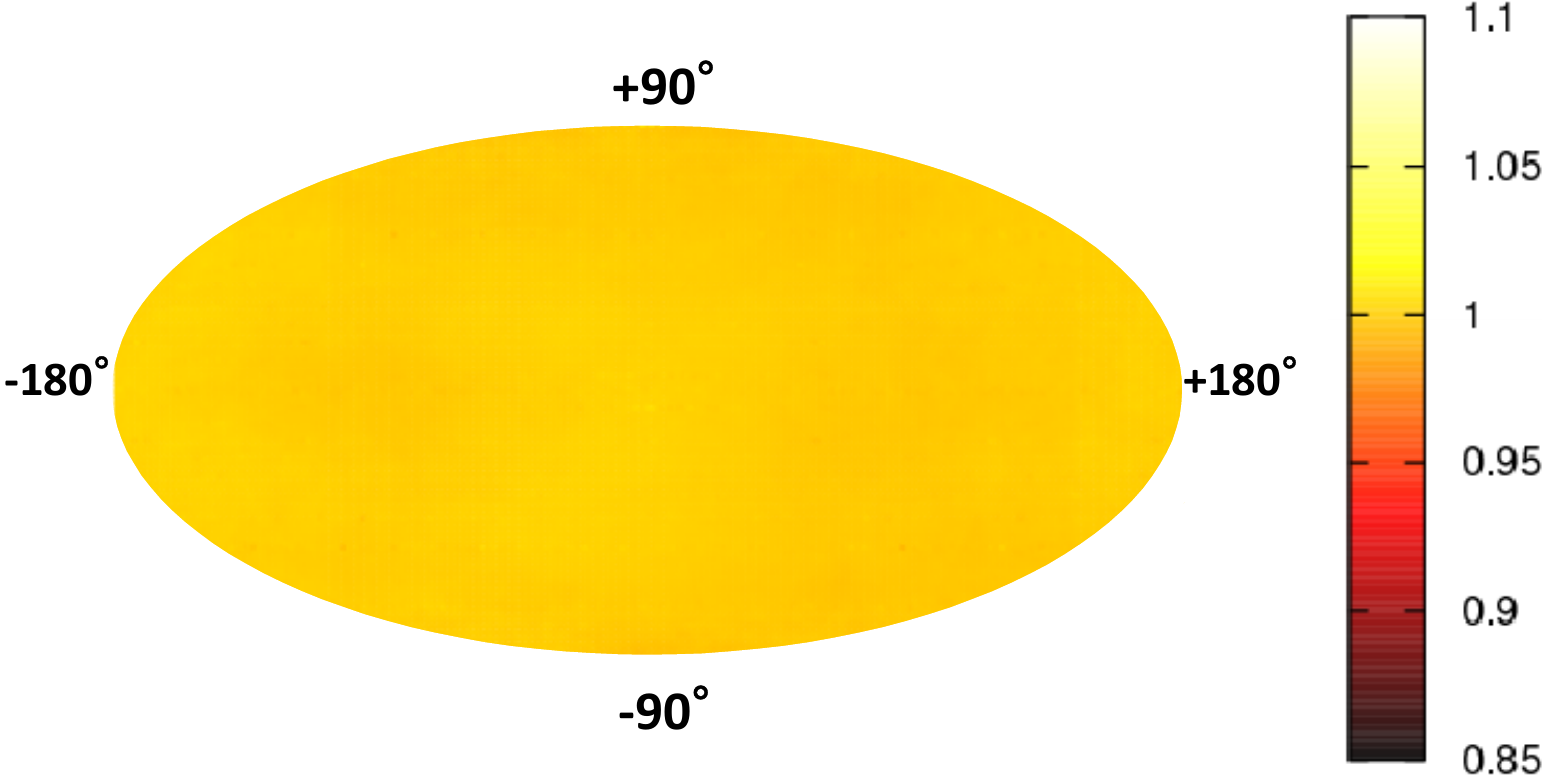}
\includegraphics[width=0.65\linewidth,bb=0 0 445 224]{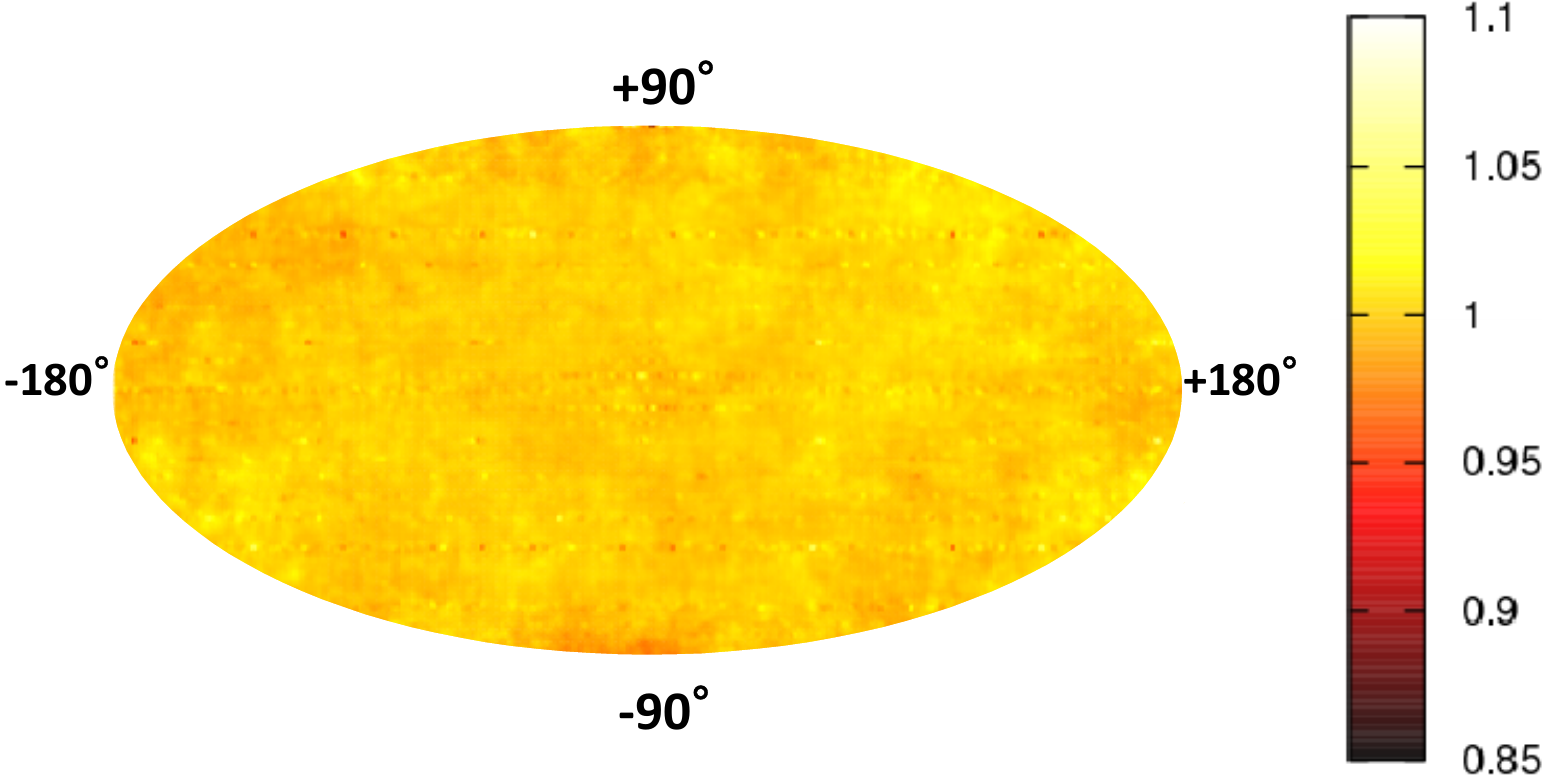}
\includegraphics[width=0.65\linewidth,bb=0 0 445 224]{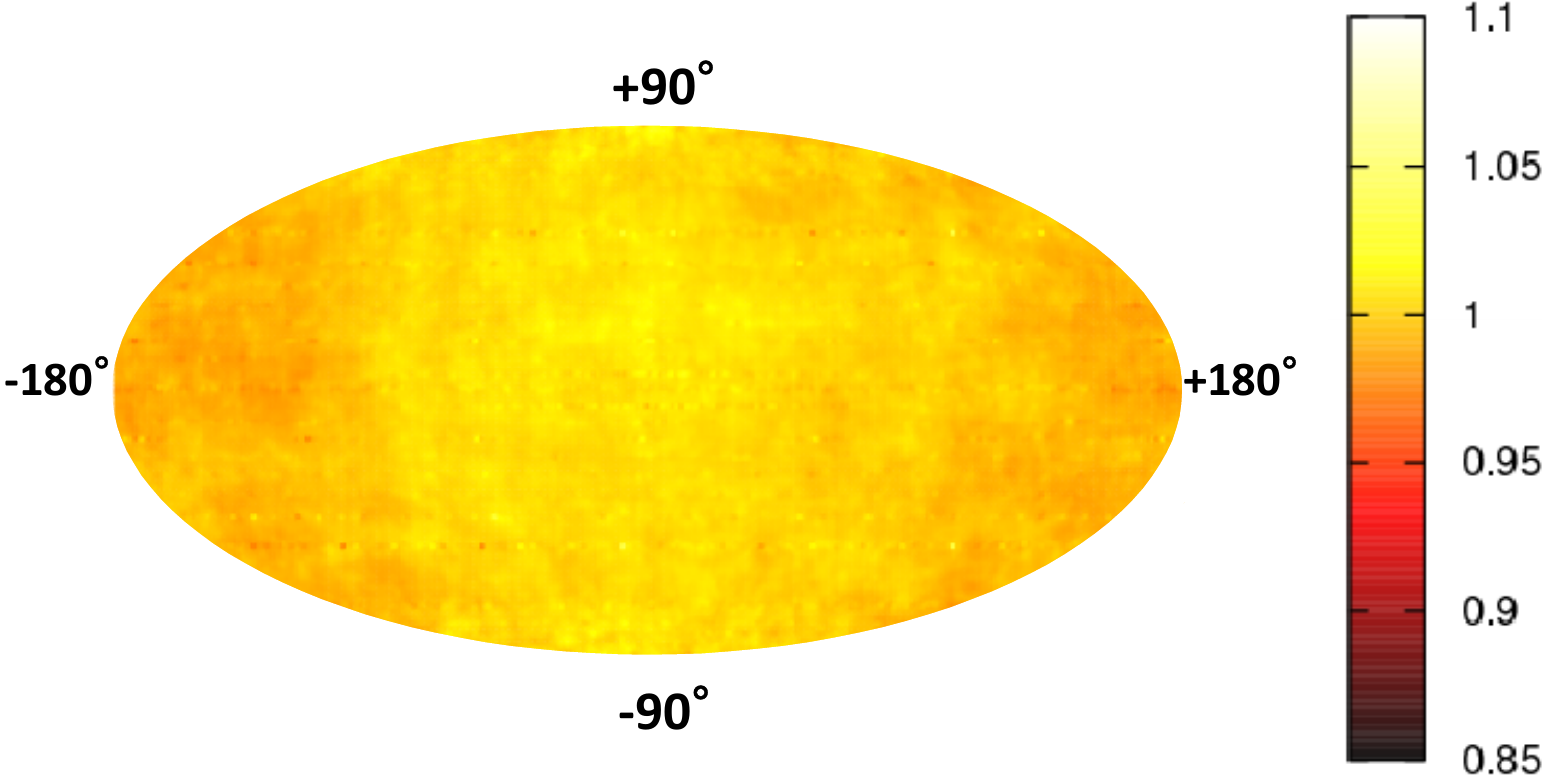}	
\end{center}
	\caption{\footnotesize 
	From the top to the bottom, 
	the angular distribution maps for 
	{\it Crowded}, 
	{\it Synchronized} and {\it Spread} cases. 
	In all cases, the anisotropy is not observed at all. 
	 }
\label{fig:fg_case1}
\end{figure}

In the above setting of the problem, 
we attempt to evaluate the $\gamma$-value using 
the same way as our previous study \cite{Makiguchi}. 
We control the weights 
of three interactions in the BOIDS 
$\mbox{\boldmath $J$}=(J_{1},J_{2},J_{3})$ 
to generate typical three cases, 
namely, 
{\it Crowded} for $(J_{1},J_{2},J_{3})=(1,0,0)$, 
{\it Synchronized} for $(J_{1},J_{2},J_{3})=(0,1,0)$ and 
{\it Spread } for $(J_{1},J_{2},J_{3})=(0,0,1)$. 
We list the $\gamma$-values and the corresponding frequency of collisions (FC)
 in Table \ref{tb:result}. 
\begin{table}[!t]
\begin{center}
\begin{tabular}{|l||c|c|c|}
	\hline
	Behaviour & Weight vector $\mbox{\boldmath $J$}$ & $\gamma$-value [$SD$] & $FC$ \\
	\hline
	{\it Crowded} & (1,0,0) & 0.332 [0.0816] & 100\% \\
	{\it Synchronized} & (0,1,0) & 0.319 [0.292] & 17.93\% \\
	{\it Spread} & (0,0,1) & 0.347 [0.304] & 0\% \\
	\hline
\end{tabular}
\end{center}
\caption{\footnotesize
Resulting $\gamma$-value and 
the {\it SD} (Standard Deviation),  
and the {\it FC} (Frequency of Collisions) 
are shown for three different ad-hoc choices of the weights 
$(J_{1},J_{2},J_{3})$. }
\label{tb:result}
\end{table}
We also show the angular distribution maps in Fig. \ref{fig:fg_case1}. 
From these table and figure, we clearly find that 
in all cases, the anisotropy is not observed at all. 
\mbox{}

\begin{figure}[!t]
 	\begin{center}
		\includegraphics[width=0.9\linewidth,bb=0 0 561 406]{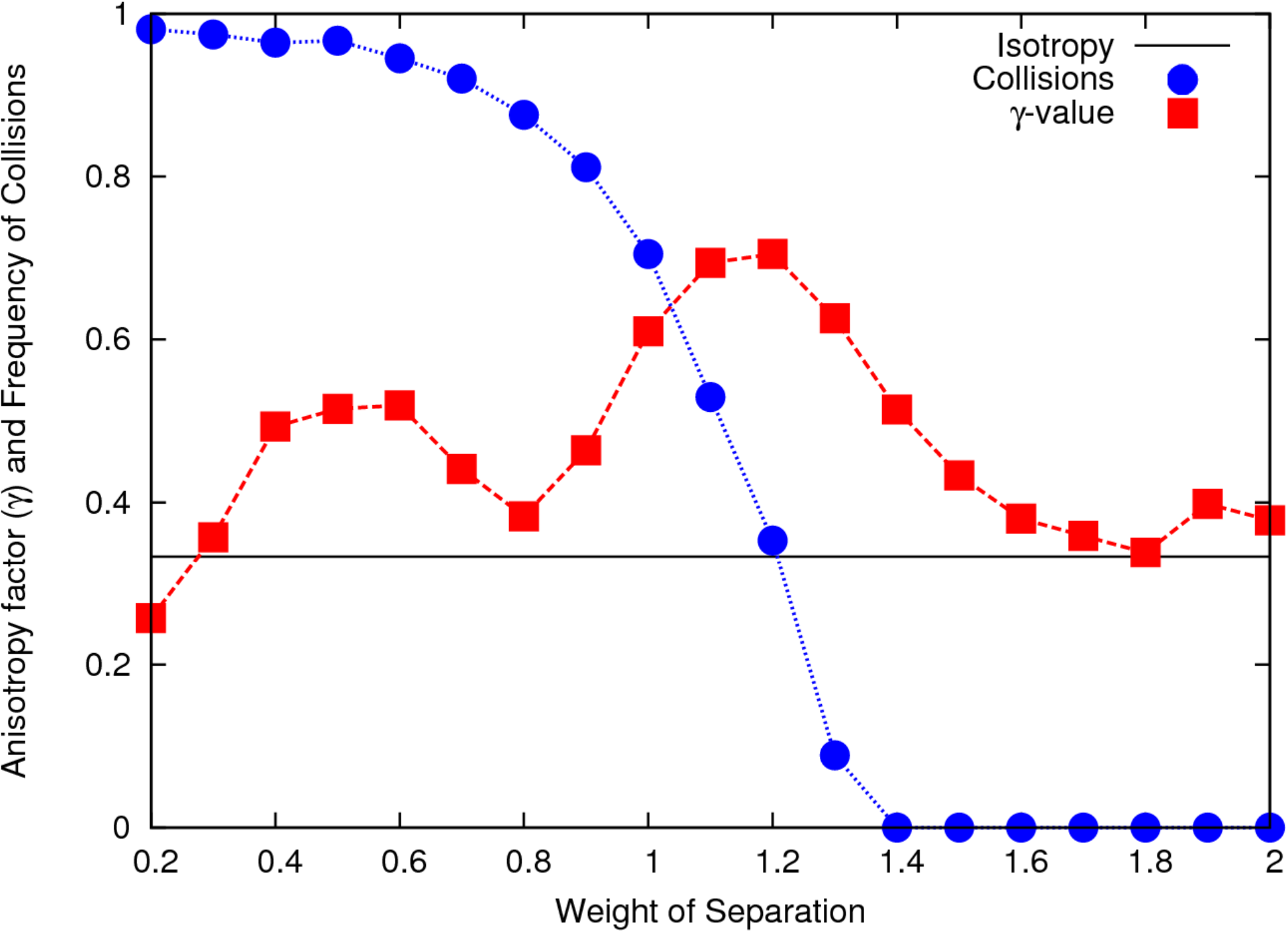}
	\end{center}
	\caption{\footnotesize 
	The $\gamma$-value and the corresponding 
	frequency of collisions as a function of 
	$J_{3}$ we set $J_{1}=1, J_{2}=5$. 
	The frequency of collisions drops to zero around $J_{3}=1.4$. 
	This fact makes us determine to choose the appropriate set as 
	$(J_{1},J_{2},J_{3})=(1,5,1.4)$.
	 }
\label{fig:fg_case2}
\end{figure}

We next choose the weights $J_{1}$ and $J_{2}$ as 
$J_{1}=1, J_{2}=5$ which we used in the previous study 
\cite{Makiguchi} as an appropriate choice to produce the 
anisotropy, and we shall vary the $J_{3}$ from $0.2$ to $2.0$
to  evaluate the $\gamma$-value and the corresponding 
frequency of collisions as a function of $J_{3}$. 
We plot them in Fig. \ref{fig:fg_case2}. 
From this figure, we find that 
the frequency of collisions decreases monotonically as 
$J_{3}$ increases and it 
converges to zero around $J_{3}=1.4$. 
On the other hand, 
the $\gamma$-value 
takes its maximum $0.70$ around 
$J_{3}=1.2$. 
From these observations, we conclude that 
we should choose the weight 
as
$\mbox{\boldmath $J$}=(J_{1},J_{2},J_{3})=(1,5,1.4)$
and whose $\gamma$-value will be about 0.70 . 
\section{Optimization to design artificial flockings}
In the previous section, we 
show on what condition the anisotropy 
emerges by setting the weights for three essential 
interactions by ad-hoc manner as a preliminary. 
Here we mention that the procedure to 
determine the interactions can be regarded as 
a kind of {\it optimization problems}. 
\subsection{Optimization under two essential constraints}
In real flockings, 
it might be a very serious problem for each agent how 
to refuse collisions with the other mates during 
the flock is moving. 
In addition, it is very hard for us to say that 
the flock splitting into several sub-flocks is also `realistic' flocking. 
Therefore, we should design the BOIDS simulations 
so as to avoid these two unexpected accidents, 
namely, `collision' and `breaking-up'. 
To realize the simulation 
in which there are no collision and breaking-up, 
we introduce two constrains into the optimization problem. 

We first define the `collision' 
as the case in which the distance between an arbitrary 
agent $i$ and its $1$-st nearest neighbouring
 mate $\overline{j(1:i)}$, say, $r_{i \overline{j(1:i)}}$ is shorter than 
their body length {\it BL}$=0.2$. 
We also assume that the `breaking-up' occurs when  
the distance between the 
center of mass and the most far agent from it becomes  
longer than a given constant. 
It is naturally imagined that 
taking into account the `collision' and 
making the algorithm to avoid it are 
essential issues not only for artificial flockings 
but also for flockings in the real world. 
It should be noted that 
the conventional 
flocking simulations based on 
the `particle models' (see for example \cite{Olfati}) in which the size of the mate is neglected 
cannot deal with the `collision'. 
We are also confirmed that avoiding the `breaking-up' also might be an 
essential factor to decide the size of the flocking.

Thus, we might use the $\gamma$-value as 
a cost function (energy function) to be minimized 
under `zero-collision' and `zero-breaking-up' constraint 
to determine the three essential interactions $\mbox{\boldmath $J$}=(J_{1},J_{2},J_{3})$. 
Namely, we should solve the following 
optimization problem with the cost. 
\begin{eqnarray}
E(\mbox{\boldmath $J$}) & = & 
\gamma (\mbox{\boldmath $J$}) + 
\lambda_{1}\mathcal{N}(\mbox{\boldmath $J$}) 
+ \lambda_{2} \mathcal{B}(\mbox{\boldmath $J$}),\,\,\,
\lambda_{1},\lambda_{2} \to \infty
\end{eqnarray}
where we defined 
$\mathcal{N}$ and $\mathcal{B}$ as the 
number of the collisions and breaking-up, respectively. 
The $\lambda_{1},\lambda_{2}$ stand for the Lagrange multipliers. 
In other words, the optimal interactions 
$\mbox{\boldmath $J$}_{\rm opt}$ is given by 
\begin{eqnarray}
\mbox{\boldmath $J$}_{\rm opt}={\rm argmax}_{\mbox{\boldmath $J$}} 
\lim_{\lambda_{1},\lambda_{2} \to \infty} 
E(\mbox{\boldmath $J$}). 
\end{eqnarray}
Since Reynolds proposed the BOIDS, 
quite a lot of the modifications or the variants were 
constructed in terms of engineering, however, 
no studies concerning the systematic determination 
of the essential three interactions in the algorithm 
from the view point of empirical observation on 
real flockings such as starlings. 
Therefore, here we formulate the procedure to 
determine the interactions as optimization 
problems having the $\gamma$-value as the cost 
under two essential constraints. 
To solve the optimization problem by means of 
the conventional tools, say, 
genetic algorithm (GA for short), 
we might design the BOIDS more systematically. 
In the next section, we shall examine the 
GA to solve our optimization 
problem to determine the optimal set of the interactions in the BOIDS. 
\section{Genetic algorithms}
Here we apply the GA to the determination of 
the weights of the interactions 
$\mbox{\boldmath $J$}$ 
in BOIDS simulations. 
Before we show the results, we 
shall briefly explain the motivation to use the GA and 
the outline of the set-up and the procedure. 
The details of the GA shall be explained in Appendix \ref{app:A}. 
\subsection{Why do we use the GA?}
The GA is a stochastic method to obtain a candidate of  
the solution having the highest possible fitness 
in the complicated fitness function with multi-valley structures. 
In GAs, 
one codes the candidates of the solution 
by a set of vectors,  each of which is 
referred to  as a `gene configuration' (a genetic code). 
Then, we make several operations, 
namely, 
{\it Crossover}, 
{\it Mutation} and {\it Selection} 
to regenerate gene configurations 
having relatively high fitness values \cite{Goldberg}. 

As a study to determine the 
weights for the interactions in the BOIDS, 
Chen {\it et. al.} \cite{Chen} 
proposed the so-called {\it Interactive genetic algorithm (IGA)}. 
However, 
we should mention here that 
they used the fitness function 
which is constructed 
subjectively, 
and in this sense, 
their approach is 
essentially different from ours. 
This is because as we already mentioned, 
we use the $\gamma$-value 
which is a measurement 
introduced by empirical findings \cite{Ballerini}. 

Of  course, there are a lot of optimization 
methods and we do not have to use 
the GA to obtain the solution to maximize the $\gamma$-value. 
However, we might assume that 
the agent (bird for instance) acquired 
such an intelligent way to behave as `flock' during their 
process of evolution and 
this assumption makes us use the GA. 
The justification of using the GA is 
very difficult to show and it might be impossible 
to prove the validity of the above assumption theoretically. 
Nevertheless, here we use the GA as a first attempt to 
design the optimal BOIDS 
based on the maximization principle of 
the $\gamma$-value as a fitness function. 
\subsection{Procedure of the GA}
In this subsection, we shall explain 
the outline of the procedure of the GA. 
In our GA, 
we use the three weights of the BOIDS, namely, 
$\mbox{\boldmath $J$}=(J_{1},J_{2},J_{3})$ as gene configurations. 
Each of the components $J_{i}$ denotes a `chromosome' 
or simply `gene' and 
takes the value in the range $[0.001, 0.999]$ and 
the minimum value of 
changing the state is set to $0.001$, 
namely, we here vary  the 
value of each component by $J_{i} \to J_{i} \pm 0.001, \,\,
i=1,2,3$ 
for each step of three operations we mentioned above. 

To operate the {\it Selection}, we need the value of 
the fitness function, 
namely, the $\gamma$-value for the 
nearest neighbouring agent 
($\gamma$-value defined by  (\ref{eq:timeAve}) 
with $n=1$). 
To evaluate it, we 
use the time-averaged $\gamma$-value 
$\mathbb{E}_{t}[\gamma_{t}]$ 
(see (\ref{eq:timeAve}) for its definition)
which is calculated by sampling
positions of the mates  every $0.1$ [sec] during $8$ [sec] 
($T=80$ data points are needed to evaluate the 
$\gamma$-value for each update of the gene configurations).  
The details of the total procedure of the GA is given in Appendix \ref{app:A}. 
 
 We also explain a border-bias free (the `border-bias'  was already mentioned in the section of 
 introduction as {\bf Bottleneck 3}) procedure to evaluate the $\gamma$-value 
in the computer simulations in Appendix \ref{app:B}. 
\section{Results}
\label{sec:result}
In Table.\ref{tbl:GA_result}, 
We show the highest $\gamma$-values 
for three independent runs, which are referred to as 
{\it Case 1,2} and {\it Case 3}, and corresponding weights for the interactions. 
It should be noted that we normalized 
the weights $\mbox{\boldmath $J$}=(J_{1},J_{2},J_{3})$ so as to 
make the maximum $J_{i}$ among the three $i=1,2,3$ unity. 
\begin{table}[ht]
\begin{center}
\begin{tabular}{|c||c||c c c|}
	\hline
	Case & $\gamma$-value & $J_1$ & $J_2$ & $J_3$\\
	\hline
	{\it Case 1} & 0.795 & 0.270 & 0.640 & 1 \\
	{\it Case 2} & 0.797 & 0.234 & 0.699 & 1 \\
	{\it Case 3} & 0.797 & 0.190 & 0.895 & 1 \\
	\hline
\end{tabular}
\end{center}
\caption{\footnotesize 
Resultant sets of weights for three interactions.}
\label{tbl:GA_result}
\end{table}
For each run, the highest $\gamma$-value is 
larger than $0.79$ and 
corresponding weights $\mbox{\boldmath $J$}=(J_{1},J_{2},J_{3})$ take 
similar values for all cases. 
In following, we 
investigate the optimization process for {\it Case 3}. 
\subsection{Optimization process of GA}
\begin{figure}[!t]
	\centering 
	\includegraphics[width=0.9\linewidth,bb=0 0 201 144]{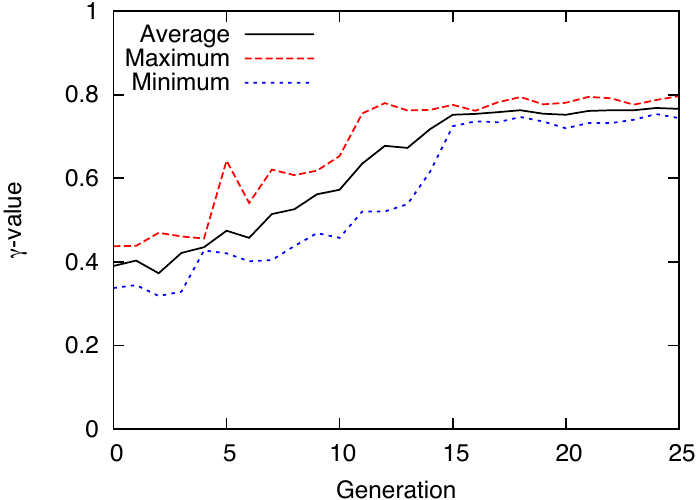}
	\caption{\footnotesize 
	Evolution of $\gamma$-value in generations.}
	\label{fig:gamma_evolution}
\end{figure}

\begin{figure}[!t]
	\centering
	\includegraphics[width=0.9\linewidth,bb=0 0 201 144]{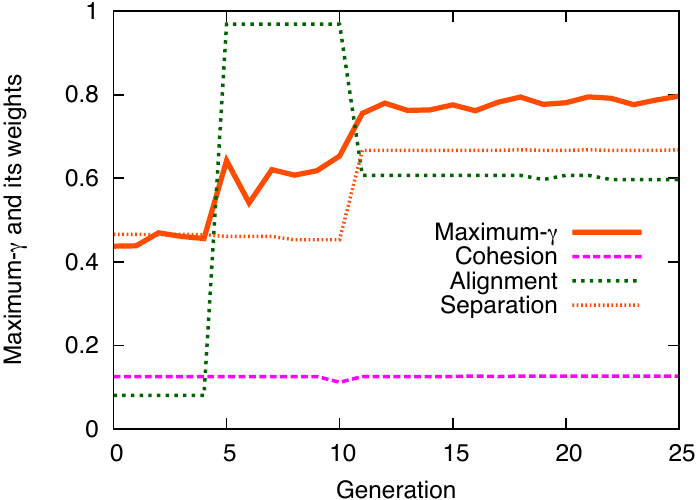}
	\caption{\footnotesize 
	Evolutions of `Maximum-$\gamma$' 
	(the highest $\gamma$-value), $J_{1}, J_{2}$ and 
	$J_{3}$ in generations. }
	\label{fig:P_evolution}
\end{figure}

\begin{figure}[!t]
	\begin{minipage}{0.49\linewidth}
	\centering
	\includegraphics[width=\linewidth,bb=0 0 190 142]{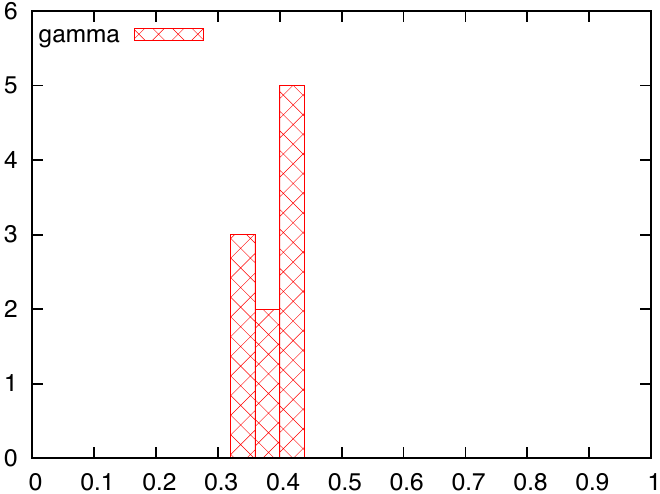}
\end{minipage}
\begin{minipage}{0.49\linewidth}
	\centering
	\includegraphics[width=\linewidth,bb=0 0 190 142]{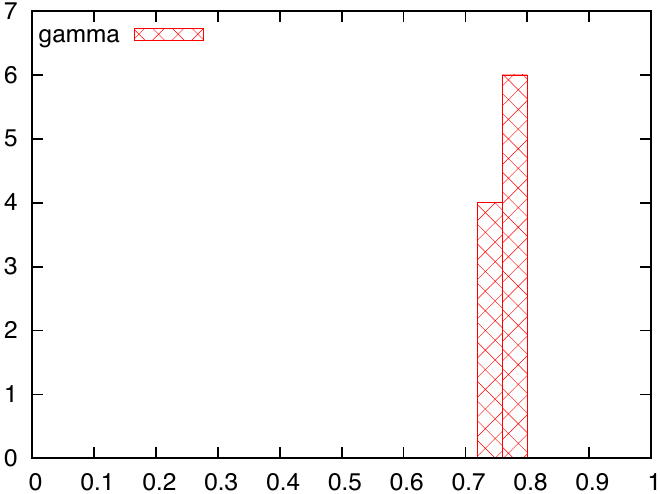}
\end{minipage}
	\caption{\footnotesize 
	{\bf Left:}  Histogram of $\gamma$-value for  the initial. 
	{\bf Right:} Histogram of $\gamma$-value for  the final. 
	We are confirmed that the GA successfully finds the value which is close to 
	the highest possible $\gamma$-value with a high probability.  
}
	\label{fig:hist_gamma}
\end{figure}

We show the minimum, average and maximum 
of the $\gamma$-value for each generation 
in Fig. \ref{fig:gamma_evolution}. 
From this figure, we find that 
these all values converge to 
$\gamma \simeq 0.8$ from the initial 
state ($\gamma \simeq 0.4$). 
We next show the time 
evolution of the weights $\mbox{\boldmath $J$}=(J_{1},J_{2},J_{3})$ for 
the best possible gene configuration 
having the highest $\gamma$-value in Fig.\ref{fig:P_evolution}. 
From this figure, we find that 
each weight changes its state during 
the GA dynamics and 
this result tells us that 
optimal gene configuration can be 
successfully generated by our GA procedure. 
We also plot the distribution of 
$\gamma$-value at the initial generation 
(Fig.\ref{fig:hist_gamma} (left)) 
and at the final generation (Fig.\ref{fig:hist_gamma} (right)). 
From these two panels, we are confirmed that 
the gene configurations having 
relatively high fitness values are generated and they actually survive 
until the final generation.  
However, 
the final distribution has a finite deviation instead of 
a single delta peak. 
This means that we could not find the optimal 
gene configuration with probability $1$. 
\begin{figure}[!t]
\centering
	\begin{minipage}{0.32\linewidth}
	\centering
	\includegraphics[width=\linewidth,bb=0 0 189 140]{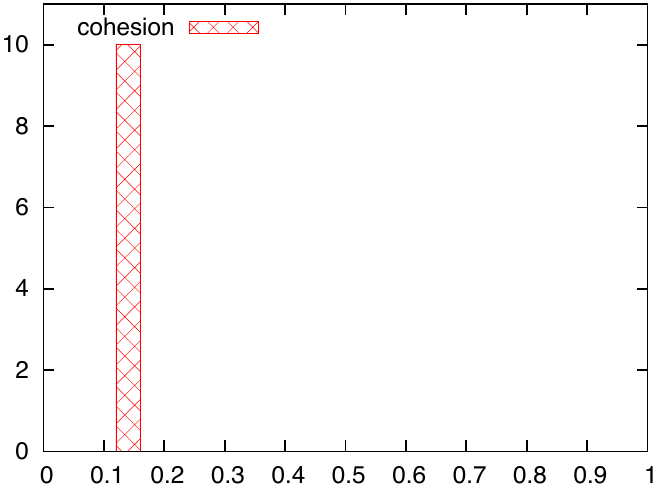}
	\end{minipage}
	\begin{minipage}{0.32\linewidth}
	\centering
	\includegraphics[width=\linewidth,bb=0 0 190 142]{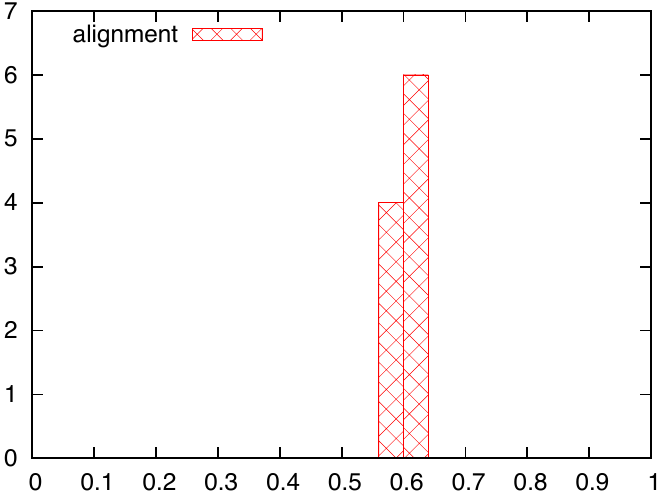}
	\end{minipage}
	\begin{minipage}{0.32\linewidth}
	\centering
	\includegraphics[width=\linewidth,bb=0 0 190 142]{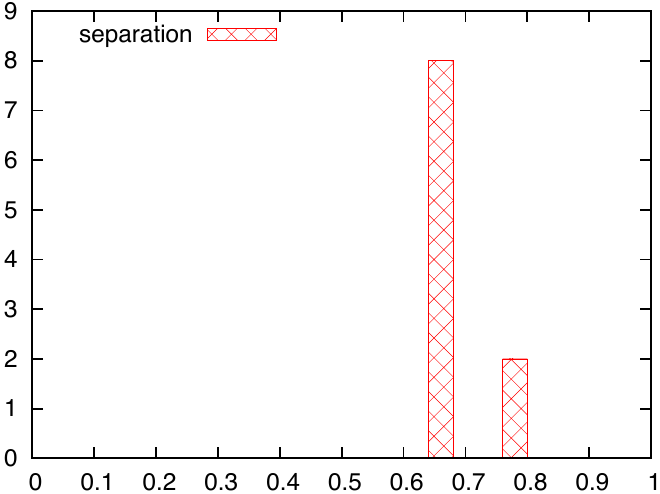}
	\end{minipage}
\caption{\footnotesize 
	Histogram of weight of each interaction for the final gene set.
	{\bf Left:} weight of `Cohesion'.
	{\bf Center:} weight of `Alignment'.
	{\bf Right:} weight of `Separation'.
	}
	\label{fig:hist_fin}
\end{figure}
\subsection{The $\gamma$-value for the optimal BOIDS}
\begin{figure}[t!]
\centering
	\includegraphics[width=0.9\linewidth,bb=0 0 509 364]{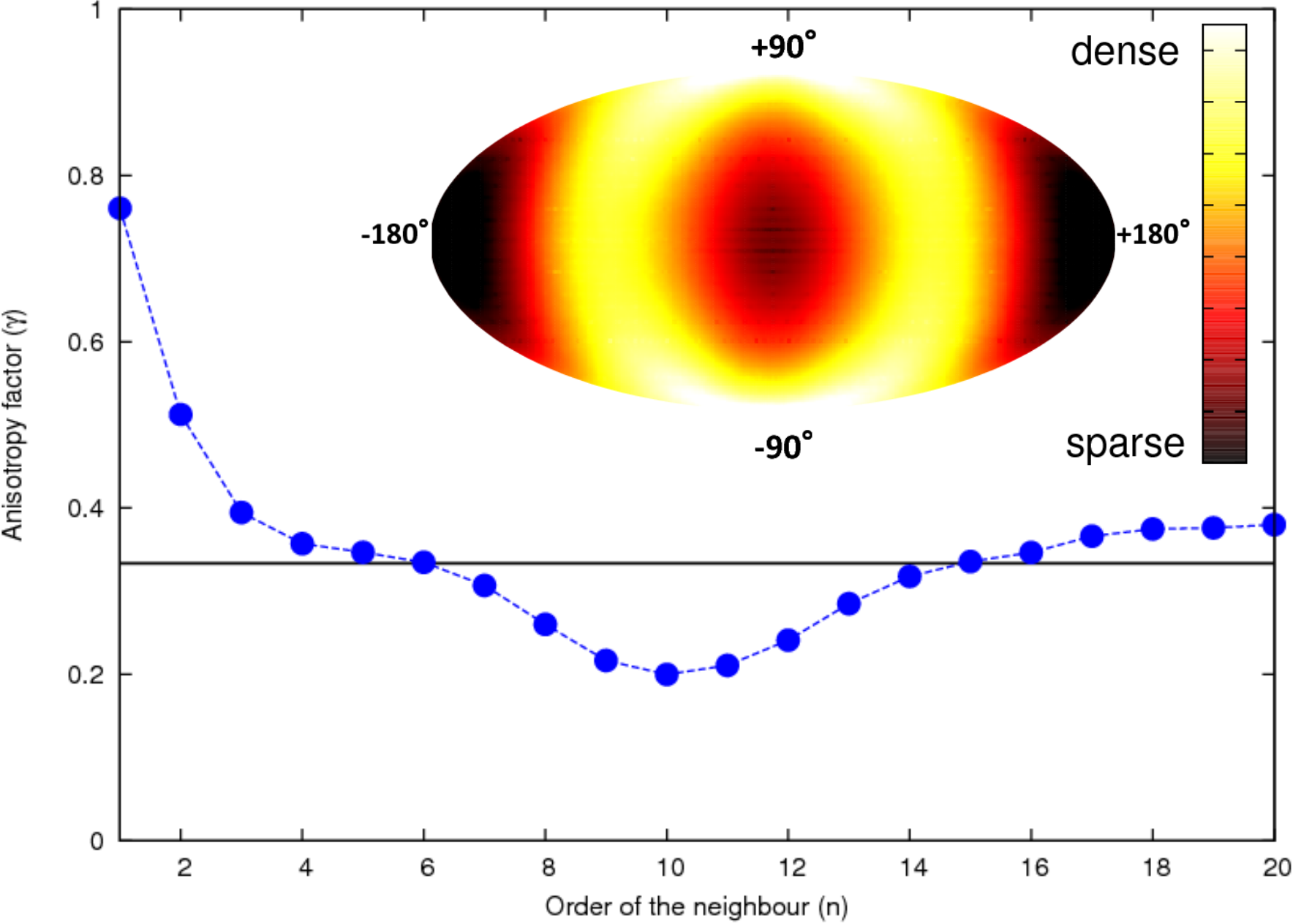}
\caption{\footnotesize 
	The $\gamma$-value as a function of $n$ calculated by the `optimal'  BOIDS simulation.
	The inset stands for the corresponding angular distribution.}
\label{fig:N_Gamma_Result}
\end{figure}

Here we examine the $n$-th nearest neighbouring agent's $\gamma$-value 
for the optimal BOIDS having the optimal weights obtained in the previous section. 
We carry out $1000$-trials to evaluate the $\gamma$-value 
for each $n$ from $n=1$ up to $n=25$. 
We also show the angular distribution map for $n=1$ and 
check the behaviour by graphical user interface (GUI).  
The results are shown in 
Fig.\ref{fig:N_Gamma_Result}. 
From this figure, we find that 
the $\gamma$-value for $n=1$ takes 
the highest value which was also observed in the empirical 
data analysis \cite{Ballerini}. 
We also find from the GUI that 
a realistic flocking's behaviour in which 
the distance between nearest neighbouring agents 
is not zero (`zero-collision') but finite 
is achieved for the BOIDS with optimal weights of the interactions . 
\subsection{Anti-anisotropy effect and its possible explanation}
In Fig.\ref{fig:N_Gamma_Result}, 
we find that the $\gamma$-value becomes lower than the 
isotropic limit $\gamma=1/3$ for $7 \leq n \leq 14$. 
This `anti-anisotropy' effect can be explained from the view point of 
geometric structure of the flocking as follows. 
\begin{figure}
\begin{center}
\includegraphics[width=0.9\linewidth,bb=84 133 746 519]{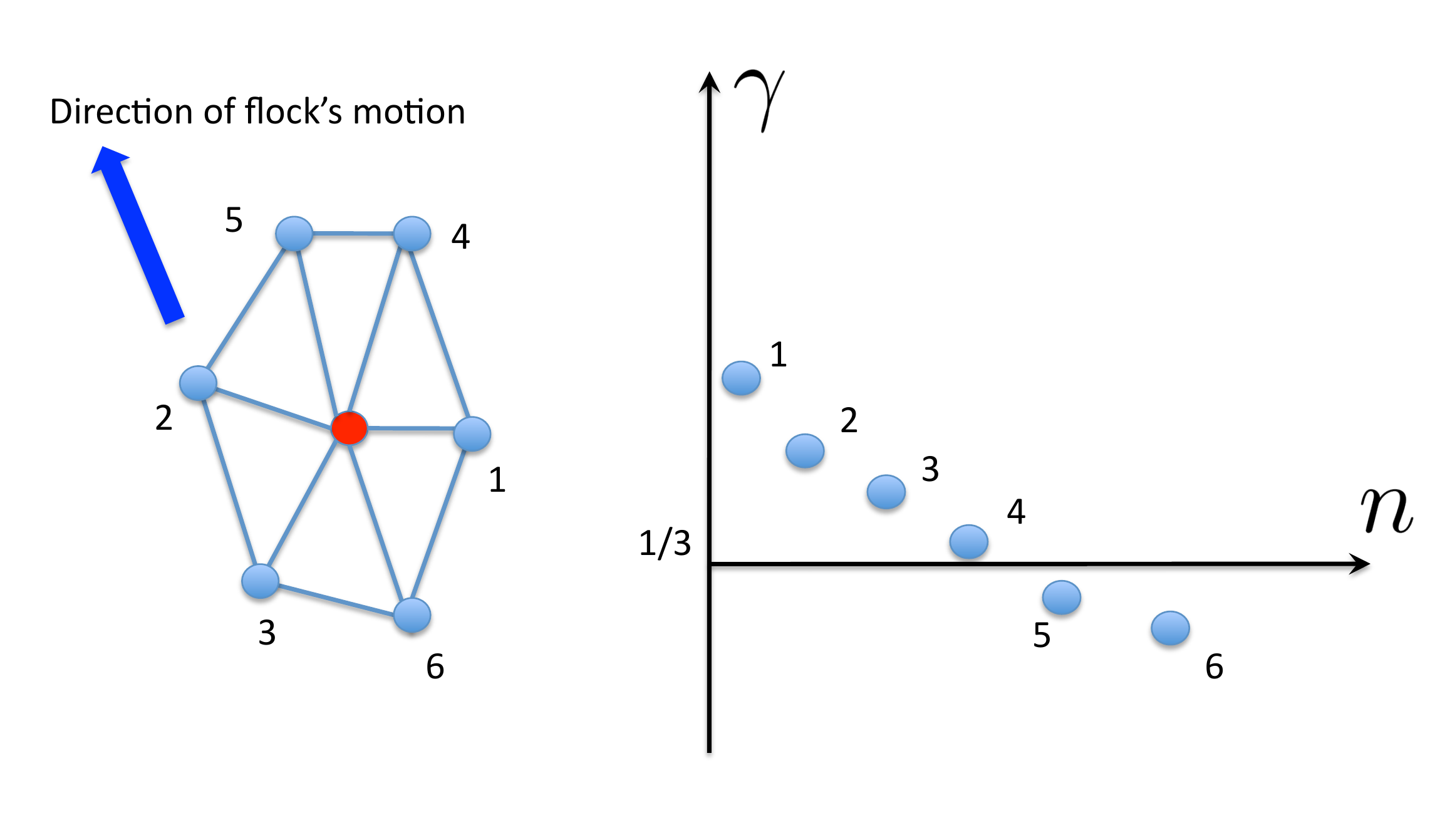}\mbox{}\vspace{4cm}\\
\end{center}
\caption{\footnotesize  
A reasonable explanation of the `anti-anisotropy'.
The fifth and the sixth nearest neighbouring 
mates are more likely to exist in the direction of flock's motion (left).  
As the result, the $\gamma$-values for the $5, 6$-th orders 
of the nearest neighbouring become lower than 
the isotropic limit $1/3$ (right). 
}
\label{fig:fg_anti}
\end{figure}
When we assume $2$-Dimensional Field and an arbitrary agent 
is surrounded by the other six mates as shown in Fig. \ref{fig:fg_anti}
(this hexagon-shape is made by equilateral triangle with neighbours), 
the fifth and the sixth nearest neighbouring mates are more likely to exist in the 
direction of flock's motion (they are indicated by `5' and `6' in Fig. \ref{fig:fg_anti} (left)).  
As the result, the $\gamma$-values for the $5,6$-th orders 
of the nearest neighbouring become lower than 
the isotropic limit $1/3$ (see Fig. \ref{fig:fg_anti} (right)). 
We should keep in mind that the `anti-anisotropy' effect might appear 
much more clearly for the flocking being longer (in the moving direction) than is wide.  

To confirm this assumption much more explicitly, 
we evaluate the third-power of average distance $R$ 
between an arbitrary agent and the $n$-th nearest neighbouring mate 
as a function of $n$.
\begin{figure}
\begin{center}
\includegraphics[width=0.9\linewidth,bb=54 249 545 575]{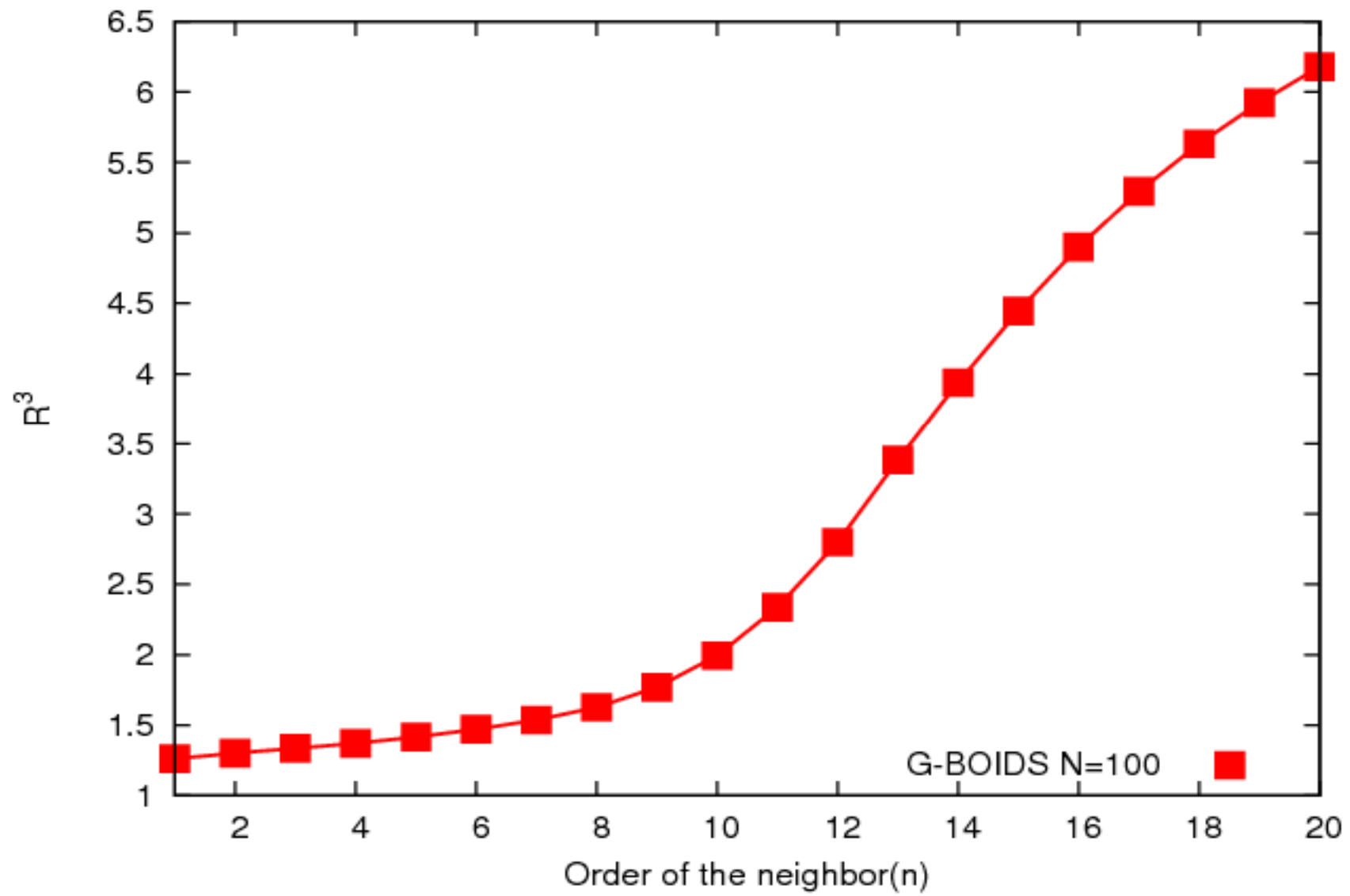}\mbox{}\vspace{6cm}
\end{center}
\mbox{}
\caption{\footnotesize 
Third-power of average distance $R$ 
between an arbitrary agent and the $n$-th nearest neighbouring mate 
as a function of order of neighbour $n$.
We are confirmed that the $R^{3}$ is almost constant 
up to the $8 \sim 10$-th nearest neighbour. 
}
\label{fig:distance}
\end{figure}
We show the result in Fig. \ref{fig:distance}. 
 From this figure, we clearly find that 
 the $R^{3}$ is almost constant up to the $8 \sim 10$-th nearest neighbour. 
This numerical result tells us that an arbitrary agent 
might be surrounded by $8 \sim 10$ mates leading to the regular-polygon structure.

In the empirical data analysis of 
starling flocking, 
such `anti-anisotropy' has never observed. 
Therefore, we might conclude that it is very hard for us to 
accept the regular-polygon structure around an arbitrary agent 
although our algorithm presented in this paper suggested the possibility
(Appendix \ref{app:A}).
\section{Topological definition of neighbours in BOIDS}
In the previous sections, we attempted to construct 
the BOIDS algorithm in which 
each agent interacts with each other when 
the distance between them is shorter than the constant 
radius of the visual field $R$. 
In this sense, we utilized the {\it metric} definition of neighbours in the BOIDS. 
In this definition of 
neighbours, the number of agents who interact with 
an arbitrary agent is not constant but apparently fluctuates. 
As we mentioned, the resulting 
$\gamma$-value shows `anti-anisotropy' due to 
the regular-polygon structure around the agent. 
Unfortunately, in real flockings, 
we have never observed such 
`anti-anisotropy' so far. 
This empirical fact tells us that 
it is less likely to exist such regular-polygon structure in the real flocks. 

In fact, Ballerini {\it et al.} \cite{Ballerini} suggested that 
a bird in the real starling flock interacts with 
a fixed number of neighbours (about six or seven neighbours). 
From this empirical findings, 
we conclude that 
the neighbours in the flocking 
should not be defined by the metric sense but it should be 
determined by the {\it topological} sense. 
Obviously, 
the topological definition of 
the neighbours is completely different from 
the metric definition which was adopted in our modelling 
of artificial flockings. 

Hence, this empirical fact also gives us 
motivations to reconsider the metric definition of 
the neighbours in the BOIDS, namely, here we assume that 
the wrong definition of the neighbours 
causes the `counter-empirical' results in 
our computer simulations. 

In this section, we shall reconstruct our BOIDS 
algorithm by taking into account the above empirical fact, 
namely, topological definition of the neighbours. 
\subsection{Topological model}
To avoid confusion, 
we first remind readers of two distinct definitions of neighbours. 

\begin{figure}[ht]
\begin{center}
\begin{minipage}{0.4\linewidth}
\mbox{}\hspace{3.5cm}
        \includegraphics[width=0.7\linewidth,bb=194 320 400 519]{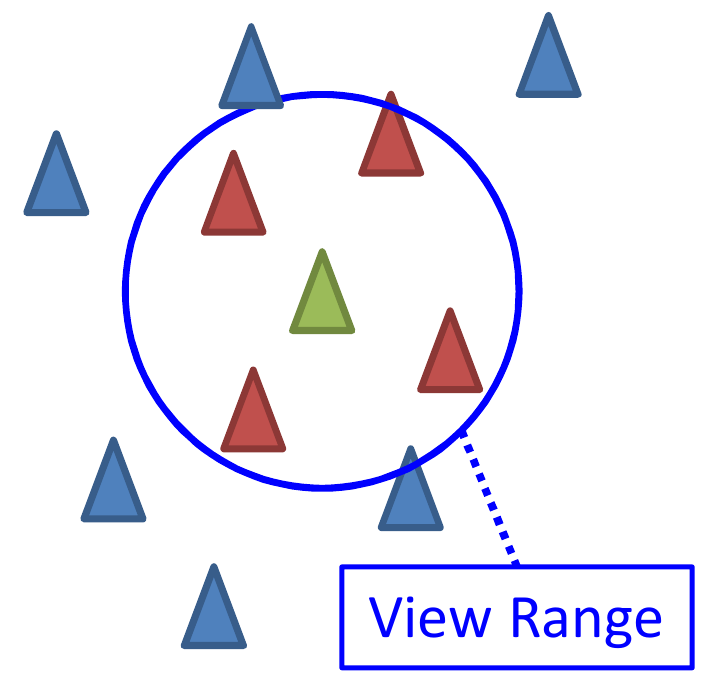} \vspace{0.5cm}
\end{minipage}
\begin{minipage}{0.4\linewidth}
\mbox{}\hspace{3.5cm}
         \includegraphics[width=0.7\linewidth,bb=187 312 406 526]{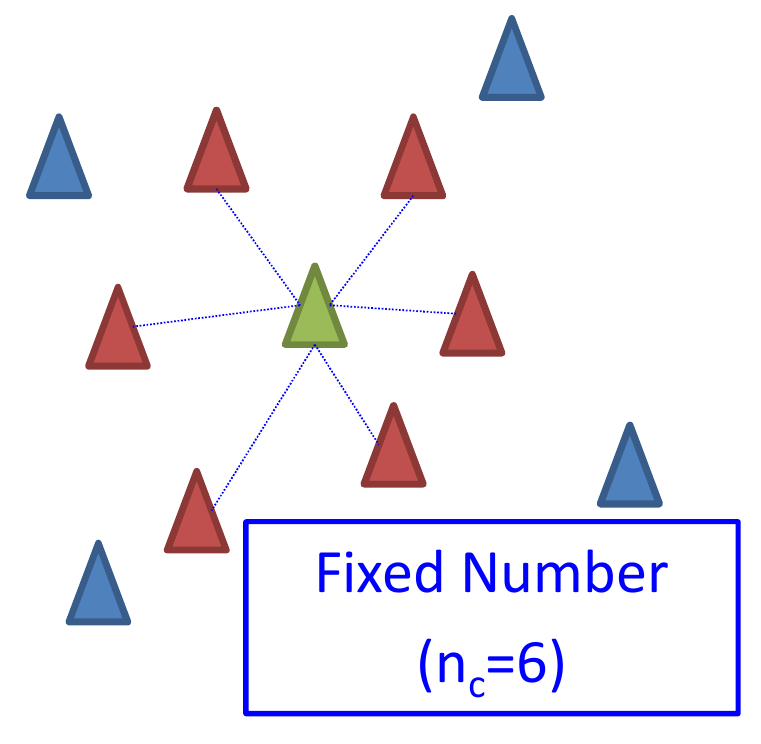} 
\end{minipage} \vspace{5cm} \\
	\caption{\footnotesize
        Two types of the definition for interacting neighbours. 
        The left panel shows the {\it metric} definition, whereas the right panel 
        corresponds to the {\it topological} definition.          
        The number of mates interacting with a given arbitrary agent is $n_{c}=6$.       
}\label{fig:def_neighbour}
\end{center}
\end{figure}

In Fig.\ref{fig:def_neighbour}, 
we show the cartoons 
for these two definitions. 
The left panel shows 
the metric definition of neighbours 
which we used in the previous sections. 
As we explained, 
each agent interacts with the others when 
the distance between mates becomes shorter than the constant 
radius of the visual field $R$. 
In the case shown in this panel, 
the agent located at the center of 
the circle interacts with four neighbours. 
On the other hand, the same agent as in 
the left panel interacts with six neighbours 
in the case of the right panel. 
The definition of the neighbours 
shown in this right panel is referred to as {\it topological}.  
Apparently, in the topological definition of neighbours, the number of mates 
interacting with a given arbitrary agent is a fixed constant and 
we define the number as $n_{c}$. 
Thus, the $n_{c}$ in the right panel of Fig. \ref{fig:def_neighbour} 
is $n_{c}=6$. 
\mbox{} 

From now on, 
the model constructed by means of the metric definition of 
neighbours is referred to as {\it metric model}, whereas we call 
the model based on the 
topological definition as {\it topological model}.

In our topological modelling, 
we set the number of interacting mates 
$n_{c}=6$, which is suggested 
by empirical data analysis by Ballerini {\it et. al.} \cite{Ballerini}. 

In order to construct the effective BOIDS simulation 
based on the topological definition of 
neighbours, we should introduce the following 
new types of interactions into our previous BOIDS. 
\begin{enumerate}
 \item[(tc)]{\bf Topological Cohesion}: Making a vector of each agent's
	    position $\mbox{\boldmath $X$}_{i}\,(i=1,\cdots,N)$ 
	    toward the average position of neighbours  
	    in the topological sense, 
	    namely, the average position over the neighbouring mates up to the $n_{c}$-th nearest neighbour.
 \item[(gc)]{\bf Global Cohesion}: Making a vector of each agent's 
	    position  $\mbox{\boldmath $X$}_{i}\,(i=1,\cdots,N)$ 
	    toward the center of mass in the flocking in order to 
	    prevent the flock from splitting into more than two distinct clusters. 
 \item[(ta)]{\bf Topological Alignment}: Keeping the velocity of each agent 
	    $\mbox{\boldmath $V$}_{i}\,(i=1,\cdots,N)$ 
	    the average value of topological neighbours (up to the
	    $n_{c}$-th nearest neighbour).
\end{enumerate}
We also slightly improve {\bf Separation} as 
\begin{enumerate}
 \item[(ms)]{\bf Modified Separation}: Making a vector of each agent 
 to avoid the collision with mates up to the $n_{c}$-th nearest neighbour in the topological sense.
\end{enumerate}
It is expected that 
the above {\bf Modified Separation} enables us 
to avoid making regular-polygon structures in artificial flockings.  
Namely, in our previous BOIDS simulations, 
the regular-polygon structure might be 
induced by metrically defined {\bf Separation} which 
acts for the only $1$-st nearest neighbour mate.
\subsection{BOIDS dynamics}
Hence, our topological model 
is described by the following update rules 
\begin{eqnarray}
\mbox{\boldmath $V$}_{i}(l+1) & = & 
\overline{V}_{l}^{(i)}
\mbox{\boldmath $e$}_{B^{'}}^{(i)}(l) 
\label{eq:BOIDS_topo} \\
\mbox{\boldmath $X$}_{i}(l+1) & = & 
\mbox{\boldmath $X$}_{i}(l) + \mbox{\boldmath $V$}_{i}(l+1)
\label{eq:BOIDS_topoX}
\end{eqnarray}
where $\mbox{\boldmath $e$}_{B^{'}}(l)$ denotes a 
unit vector pointing to 
the direction to which the agent $i$ should move 
according to the above interactions of BOIDS and explicitly given by 
\begin{eqnarray}
\mbox{\boldmath $e$}_{B^{'}}^{(i)}(l) & =& 
\frac{
\frac{J_{1}\mbox{\boldmath $v$}_{\rm TC}^{(i)} (l)+ 
J_{2} \mbox{\boldmath $v$}_{\rm TA}^{(i)}(l)
+J_{3}\mbox{\boldmath $v$}_{\rm MS}^{(i)}(l)
+J_{4}\mbox{\boldmath $v$}_{\rm GC}^{(i)}(l)}
{|
J_{1}\mbox{\boldmath $v$}_{\rm TC}^{(i)} (l)+ 
J_{2} \mbox{\boldmath $v$}_{\rm TA}^{(i)}(l)
+J_{3}\mbox{\boldmath $v$}_{\rm MS}^{(i)}(l)
+J_{4}\mbox{\boldmath $v$}_{\rm GC}^{(i)}(l)
|} 
+ \eta 
 \frac{\mbox{\boldmath $V$}_{i}(l)}{|\mbox{\boldmath $V$}_{i}(l)|}}
{
\left|
\frac{J_{1}\mbox{\boldmath $v$}_{\rm TC}^{(i)} (l)+ 
J_{2} \mbox{\boldmath $v$}_{\rm TA}^{(i)}(l)
+J_{3}\mbox{\boldmath $v$}_{\rm MS}^{(i)}(l) 
+J_{4} \mbox{\boldmath $v$}_{\rm GC}^{(i)}(l) 
}
{|
J_{1}\mbox{\boldmath $v$}_{\rm TC}^{(i)} (l)+ 
J_{2} \mbox{\boldmath $v$}_{\rm TA}^{(i)}(l)
+J_{3}\mbox{\boldmath $v$}_{\rm MS}^{(i)}(l)
+J_{4}\mbox{\boldmath $v$}_{\rm GC}^{(i)}(l)|
} 
+ \eta 
\frac{\mbox{\boldmath $V$}_{i}(l)}{|\mbox{\boldmath $V$}_{i}(l)|}
\right|} 
\label{eq:BOIDS_topo2}
\end{eqnarray}
with 
\begin{eqnarray}
\mbox{\boldmath $v$}_{\rm TC}^{(i)} (l) & = & 
\frac{
\frac{\sum_{j=1}^{N}
\Theta (R_{n_{c}}^{(i)}-r_{ij})  
\mbox{\boldmath $X$}_{j}(l)}
{
|\sum_{j =1}^{N}
\Theta (R_{n_{c}}^{(i)}-r_{ij})  
\mbox{\boldmath $X$}_{j}(l)|}
 -\mbox{\boldmath $X$}_{i}(l)}
{
\left|
\frac{
\sum_{j=1}^{N}
\Theta (R_{n_{c}}^{(i)}-r_{ij}) 
\mbox{\boldmath $X$}_{j}(l)
}
{
|\sum_{j=1}^{N}
\Theta (R_{n_{c}}^{(i)}-r_{ij}) 
\mbox{\boldmath $X$}_{j}(l)|
}
-\mbox{\boldmath $X$}_{i}(l)
\right|}  
\label{eq:VTC_topo} \\
\mbox{\boldmath $v$}_{\rm A}^{(i)}(l) & = & 
\frac{
\sum_{j=1}^{N} \Theta (R_{n_{c}}^{(i)}-r_{ij})
\mbox{\boldmath $V$}_{j}(l)
}
{|
\sum_{j =1}^{N}
\Theta (R_{n_{c}}^{(i)}-r_{ij})  
\mbox{\boldmath $V$}_{j}(l)|} 
\label{eq:VA_topo} \\
\mbox{\boldmath $v$}_{\rm S}^{(i)} (l)  & = & 
-\frac{
\sum_{j =1}^{N}
\Theta (R_{n_{c}}^{(i)}-r_{ij})
(\mbox{\boldmath $X$}_{j}(l)-
\mbox{\boldmath $X$}_{i}(l))
}
{|
\sum_{j=1}^{N} 
\Theta (R_{n_{c}}^{(i)}-r_{ij})
(\mbox{\boldmath $X$}_{j}(l)-
\mbox{\boldmath $X$}_{i}(l))|}
\label{eq:VS_topo} \\
\mbox{\boldmath $v$}_{\rm GC}^{(i)}(l) & = & 
\frac{
\frac{\sum_{j=1}^{N}
\mbox{\boldmath $X$}_{j}(l)}
{
|\sum_{j =1}^{N}
\mbox{\boldmath $X$}_{j}(l)|}
 -\mbox{\boldmath $X$}_{i}(l)}
{
\left|
\frac{
\sum_{j=1}^{N}
\mbox{\boldmath $X$}_{j}(l)
}
{
|\sum_{j=1}^{N}
\mbox{\boldmath $X$}_{j}(l)|
}
-\mbox{\boldmath $X$}_{i}(l)
\right|}  
\label{eq:VGC_topo} 
\end{eqnarray}
where 
$R_{n_{c}}^{(i)}$ 
denotes the square distance 
between the 
agent $i$ and the $n_{c}$-th nearest neighbouring mate. 
Therefore, the number $n_{c}$ should be defined explicitly by
\begin{eqnarray}
n_{c} & = & 
\sum_{j=1}^{N}
\Theta (R_{n_{c}}^{(i)}-r_{ij})
\end{eqnarray}
for all $i$ because $\Theta (\cdots)$ survives 
only for the $j$ satisfying $r_{ij} <R_{n_{c}}^{(i)}$, 
and the number of such $j$ is just $n_{c}$ from the definition. 
The balance parameter $\eta$ is set to the same value $2$ as 
in the case of the metric model. 

From the above definition of (\ref{eq:VTC_topo}), 
we easily find that 
$\mbox{\boldmath $v$}_{\rm TC}^{(i)}(l)=-
\mbox{\boldmath $v$}_{\rm MS}^{(i)}(l)$ and 
one of these two distinct effects is completely cancelled in the BOIDS 
dynamics (\ref{eq:BOIDS_topo})(\ref{eq:BOIDS_topoX}) 
as $ \sim (J_{1}-J_{3}) \mbox{\boldmath $v$}_{\rm TC}^{(i)}(l)$ 
for any choice of $J_{1},J_{3}$.
To modify this undesirable situation, 
we slightly change the 
$\mbox{\boldmath $v$}_{\rm S}^{(i)}(l)$ 
as follows. 
\begin{equation}
\mbox{\boldmath $v$}_{\rm MS}^{(i)} (l) = 
-\frac{
\sum_{n=1}^{n_{c}}
\Theta (R_{n}^{(i)}-r_{i\overline{j(n:i)}})
(\mbox{\boldmath $X$}_{\overline{j(n:i)}}(l)-
\mbox{\boldmath $X$}_{i}(l))
}
{|
\sum_{n=1}^{n_{c}}
\Theta (R_{n}^{(i)}-r_{i\overline{j(n:i)}})
(\mbox{\boldmath $X$}_{\overline{j(n:i)}}(l)-
\mbox{\boldmath $X$}_{i}(l))
|}
\label{eq:VS_topo_mod}
\end{equation}
where 
$\overline{j(i:n)}$ is given by the definition 
(\ref{eq:def_nth_nn}),
and $R_{n}$ means the Separation Range for 
the $n$-th nearest neighbour mate.

From the empirical evidence \cite{Ballerini}, 
we set $R_{n}^{(i)}$ as 
\begin{eqnarray}
R_{n}^{(i)} & = & 
|r_{0}^{(i)}| n^{1/3} 
\end{eqnarray}
where we define $R_{0}^{(i)}=R_{0}=0.73$ 
and $r_{0}^{(i)}$ 
is selected as a Gaussian variable with 
mean $R_{0}$ and unit variance.

From equation (\ref{eq:BOIDS_topo}), 
we should notice that the amplitude of 
velocity vector of agent $i$ at time $l+1$ is identical to 
the average amplitude of 
velocity vectors for the topologically defined neighbouring mates, 
that is, the mates up to the  $n_{c}$-th nearest neighbours in the previous time step $l$ as 
\begin{eqnarray}
|\mbox{\boldmath $V$}_{i}(l+1)| & = & 
\overline{V}_{l}^{(i)} \equiv 
\frac{
\sum_{n=1}^{n_{c}}
\Theta (R_{n}^{(i)}-r_{i\overline{j(i:n)}})
|\mbox{\boldmath $V$}_{\overline{j(i:n)}}(l)|}
{
\sum_{n=1}^{n_{c}}
\Theta (R_{n}^{(i)}-r_{i\overline{j(i:n)}})}.
\end{eqnarray}
The above update rules 
(\ref{eq:BOIDS_topo}),
(\ref{eq:BOIDS_topoX}),
(\ref{eq:BOIDS_topo2}), 
(\ref{eq:VTC_topo}),
(\ref{eq:VA_topo}), 
(\ref{eq:VS_topo_mod}) 
and (\ref{eq:VGC_topo}) 
are our basic 
dynamical equations to be evaluated numerically. 

We list a set of 
scale-lengths appearing in 
our simulations in Table \ref{tbl:topo_para}.  
In both Table \ref{tab:tb1} (metric model) and 
Table \ref{tbl:topo_para} (topological model), 
we chose these scale-lengths 
from the empirical data \cite{Ballerini}. 
However, 
we should keep in mind that the choices of 
$R_{0}$ are different in both cases. 
In the metric model, we used 
$R_{0}=1.09$ which is 
chosen from {\sf Event 29-03} in 
the reference \cite{Ballerini2}, 
whereas 
$R_{0}=0.73$ in the topological model comes from 
{\sf Event 28-10} in \cite{Ballerini2}. 
\begin{table}[htb]
\centering
\begin{tabular}{l|c}
\hline
\hline
\multicolumn{2}{c}{A set of scale-lengths in the topological model}\\
\hline
Number of agents ($N$) & 100\\
Body-Length ($BL$) & 0.2 [m]\\
Wing-Span ($WS$) & 0.4 [m] \\
Radius of Separation Range ($R_0$) & 0.73 [m]\\
Initial Speed Average $V^{'}$ & 11.10 [m/s]\\
Initial Density of the Aggregation ($\rho$) & 0.54 [${\rm m}^{-3}$] \\
\hline
\hline
\end{tabular}
\caption{\footnotesize
A set of scale-lengths in the topological model.
Variables other than the {\it Number of agents}
 are based on empirical data by Ballerini {\it et. al.} 
({\sf Event 28-10} in Table 1 of \cite{Ballerini2}).}
\label{tbl:topo_para}
\end{table}

For the topological model obtained by the above modifications, 
we utilize the GA to find the 
weights of the four interactions ((tc),(gc),(ta) and (ms)). 
Then, we numerically evaluate 
the $\gamma$-value and the 
the third-power of average distance ($R^3$) 
between an arbitrary agent and the $n$-th nearest neighbour
as a function of order of neighbour $n$.
\subsection{Results}
\begin{figure}[ht]
\begin{center}
\begin{minipage}{0.49\linewidth}
	 \includegraphics[width=\linewidth,bb=55 285 550 642]{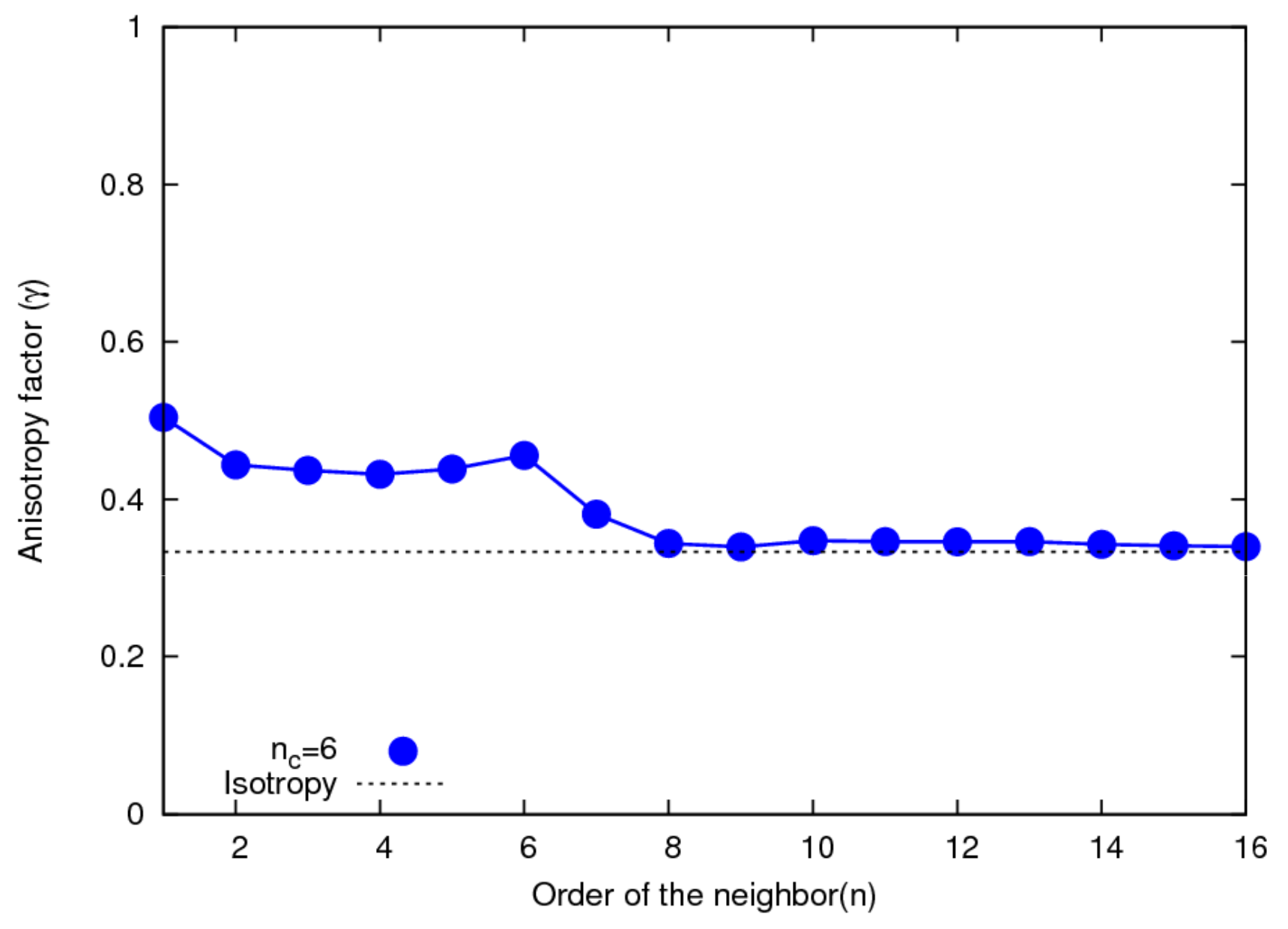} \vspace{4.5cm}
\end{minipage} 
\begin{minipage}{0.49\linewidth}
	 \includegraphics[width=\linewidth,bb=55 105 547 457]{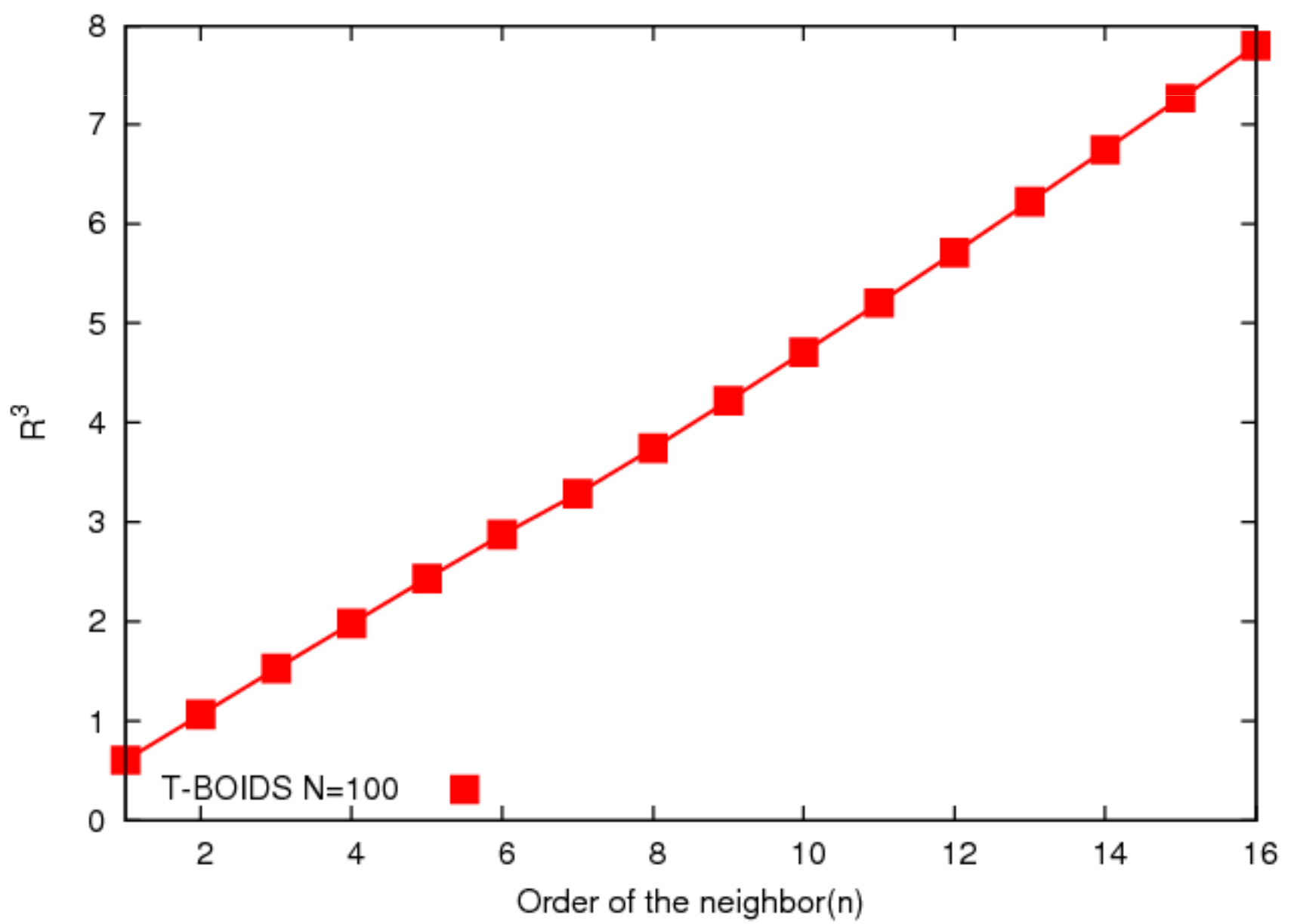}
\end{minipage} 
	\caption{\footnotesize 
        The resulting measurements for our topological model.
        The left panel shows the $\gamma$-value as a function of order $n$, 
        whereas the right panel is plotted as third-power of average distance $R$ 
between an arbitrary agent and the $n$-th nearest neighbouring mate 
as a function of order of neighbour $n$. 
We set $n_c=6$, which is indicated by empirical evidence \cite{Ballerini}.}
\label{fig:result_top}
\end{center}
\end{figure}

We show the result in Fig.\ref{fig:result_top}.
We plot the $\gamma$-value as a function of order $n$ (left panel) and 
the $R^3$ as a function of order of neighbour $n$ (right panel) for 
$n_{c}=6$. 

In the left panel, we find that the anisotropy emerges ($\gamma>1/3$) 
up to $n_{c}=6$ and `anti-anisotropy' disappears as we expected. 
From the right panel, we are also confirmed that 
the regular-polygon structures never emerges in the flock 
because the $R^{3}$ monotonically increases as the $n$ increases.  
Of course, 
here we used the empirical finding 
$R_{n}=R_{0}n^{1/3}$ \cite{Ballerini} 
to determine the $R_{n}^{(i)}$ in 
the $\mbox{\boldmath $v$}_{\rm MS}^{(i)}(l)$ for our topological modelling, 
however, the result might justify that 
the flock in the metric model behaves as a `crystal form', 
whereas the flock in the topological model 
looks like `gas'  which is much closer to real flockings. 
\mbox{}

We also calculate the 
so-called {\it integrated conditional density} $\Gamma (r)$ 
and {\it pair distribution function} $g(r)$ 
introduced by Cavagna {\it et. al.} \cite{Cavagna2008}.  
These two quantities are explicitly defined as 
\begin{eqnarray}
\Gamma (r) & = & 
\frac{1}{n_{c}}
\sum_{i=1}^{n_{c}}
\frac{N_{i}(r)}{4\pi r^{3}/3} \\
g (r) & = & 
\frac{1}{4\pi r^{2}}
\frac{1}{n_{c}}
\sum_{i=1}^{n_{c}}
\sum_{j \neq i}
\delta (r -r_{ij})
\end{eqnarray}
where $N_{i}(r)$ denotes the number of 
points in the sphere with the radius 
$r$ centered in $i$, and 
$n_{c}$ stands for the 
number of individuals in the sphere. 
$r_{ij}$ is the absolute distance 
between the center $i$ and the neighbour $j$. 
Here we should notice that 
these two quantities are related each other and satisfy 
\begin{eqnarray}
g(r) & = & 
\Gamma (r) + 
\frac{r}{3} 
\frac{d\Gamma (r)}{dr}.
\end{eqnarray}
Hence, we directly evaluated 
$\Gamma (r)$ from our simulations and 
calculated the $g(r)$ by means of the above equation. 

\begin{figure}[ht]
\begin{center}
\begin{minipage}{0.49\linewidth}
\includegraphics[width=\linewidth,bb=76 253 526 578]{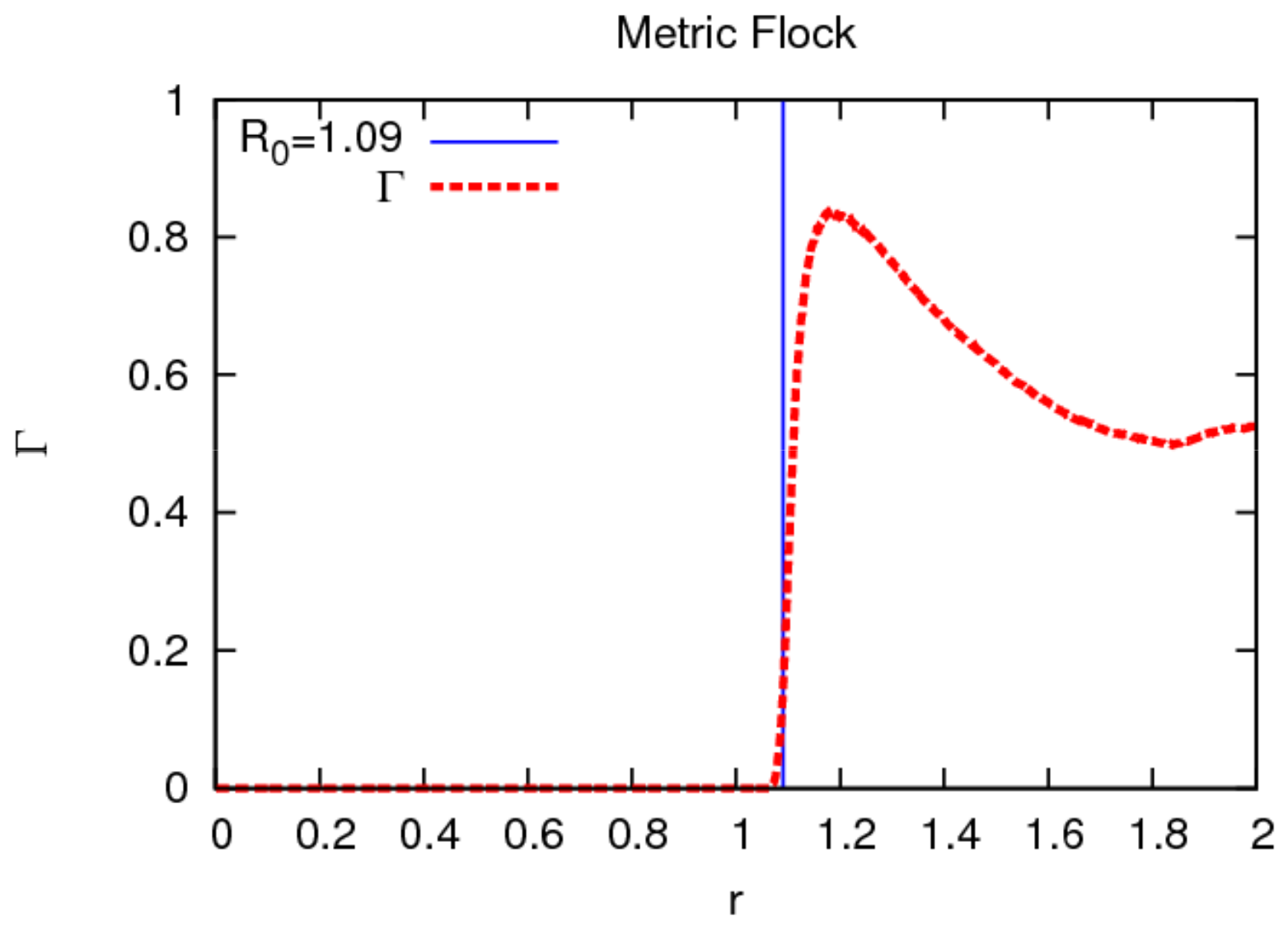} 
\end{minipage}
\begin{minipage}{0.49\linewidth}
\includegraphics[width=\linewidth,bb=76 252 528 577]{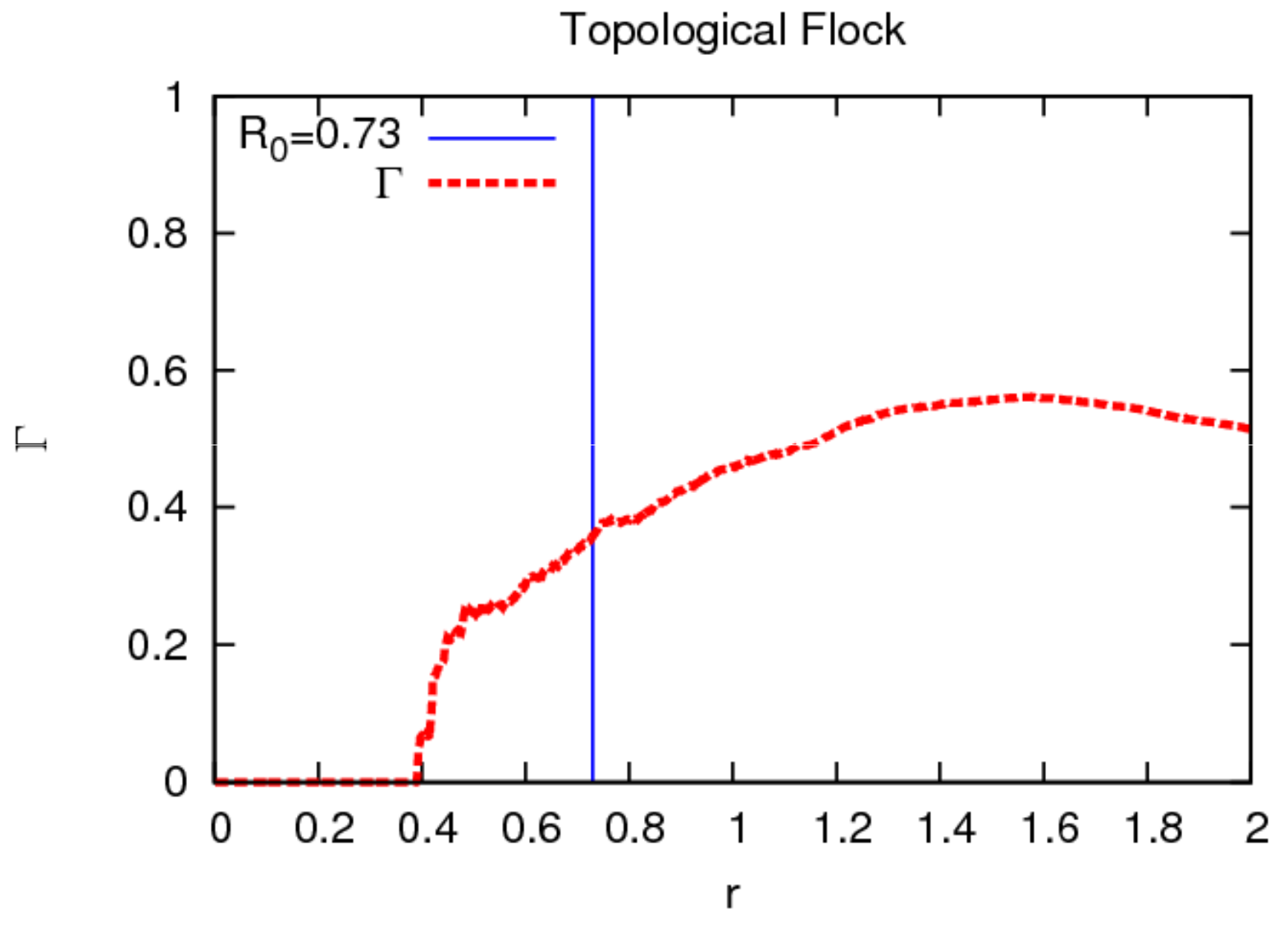} 
\end{minipage}  
\begin{minipage}{0.49\linewidth}
\mbox{}\hspace{0.5cm}
\includegraphics[width=\linewidth,bb=104 325 502 618]{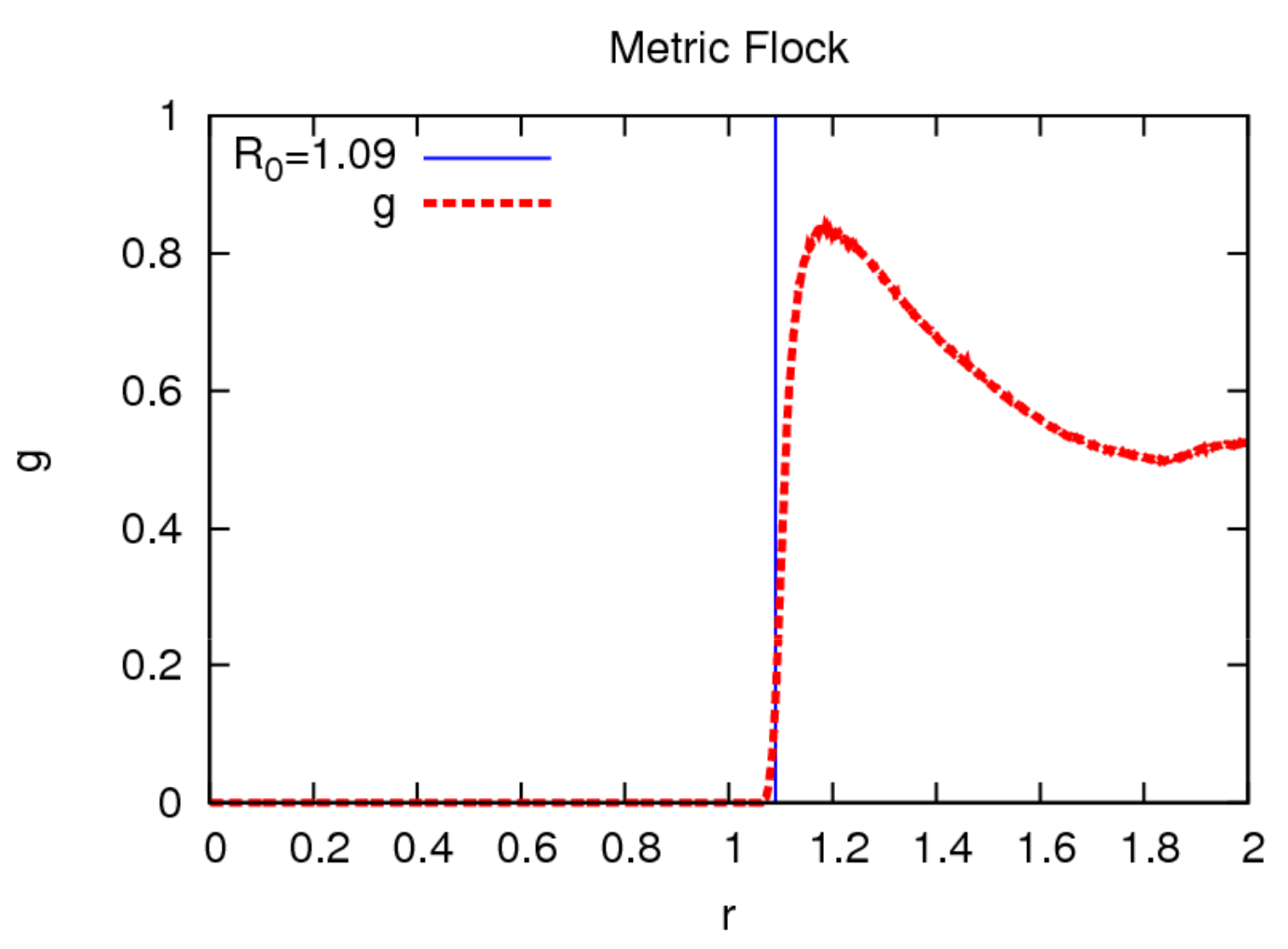} 
\end{minipage}
\begin{minipage}{0.49\linewidth}
\mbox{}\hspace{0.5cm}
\includegraphics[width=\linewidth,bb=103 324 501 611]{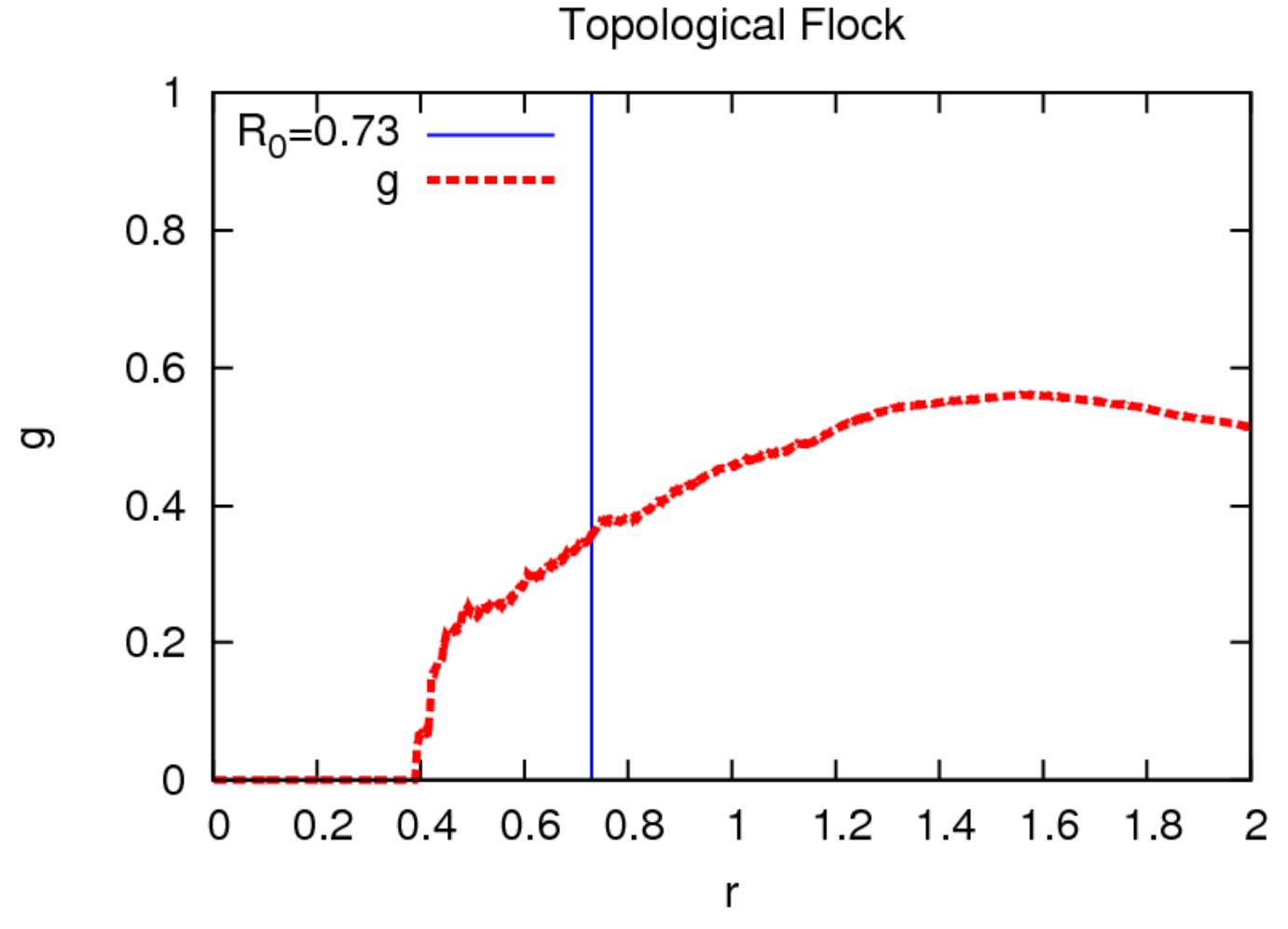}
\end{minipage}  \vspace{6cm}\\
	\caption{\footnotesize 
	 The integrated conditional density $\Gamma (r)$ (upper panels) 
	 and the pair distribution function $g(r)$ (lower panels) for 
	 the metric model (left panels) and the topological model (right panels).
	}
\label{fig:result_Gammag}
\end{center}
\end{figure}

We show the results in Fig. \ref{fig:result_Gammag}.  
From this figure, we find 
that both $\Gamma$ and $g$ 
for the metric model 
suddenly increase 
around the radius of view field $R_{0}=1.09$ and 
the behaviour is completely different from the empirical evidence \cite{Cavagna2008}. 
On the other hand, 
these quantities for the topological model gradually 
increases around $r=0.4 < R_{0}=0.73$ and 
the behaviour is very close to 
the empirical evidence 
(see Fig. 1 and Fig. 5 in \cite{Cavagna2008}). 
\mbox{}

From these numerical results, 
we conclude that our topological model reconstructs
 the essential features confirmed from empirical findings of starling.
Although some tiny gaps have been left, 
for instance, the $\gamma$-value of the nearest neighbour is lower than 
the empirical evidence, 
one can say that our topological model is more likely to be `realistic'  
than the metric model.  
\section{Discussion}
In this section, we discuss several issues  
on our results.
\subsection{On the optimal gene configuration}
From Table \ref{tbl:GA_result}, 
we find that 
the order of the strength of the 
weight $J_{1},J_{2}$ and $J_{3}$ is 
$J_{3} > J_{2}>J_{1}$ in any run of the GA.
The condition 
$J_{3}>J_{1}$ is needed to require the `zero-collision' constraint, whereas 
the condition
$J_{3}>J_{2}>J_{1}$ 
makes the $\gamma$-value larger than that for 
the condition $J_{3}>J_{1}$. 
Actually, in our previous study \cite{Makiguchi}, 
we carried out the simulations under the condition 
$J_{3}>J_{1}>J_{2}$, which was determined by hand,  
and found that the $\gamma$-value for $n=1$ is around 
$0.7$ which is apparently lower than the 
result of empirical findings $\gamma \simeq 0.8$ \cite{Ballerini}. 
This fact is a reasonable advantage of 
our approach based on the GA to maximize the 
$\gamma$-value. 
\begin{figure}
	\centering 
	\includegraphics[width=0.9\linewidth,bb=27 166 567 677]{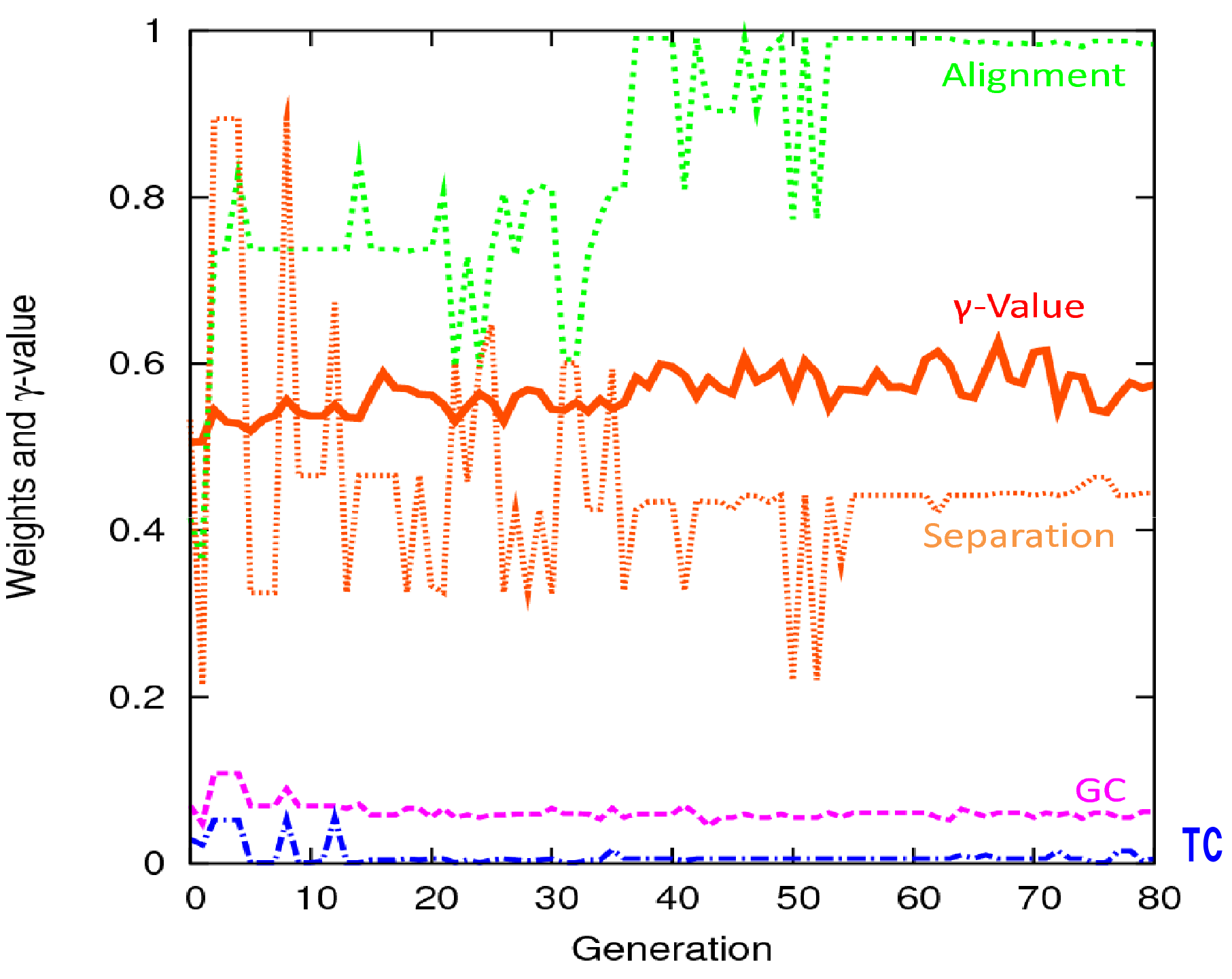} \vspace{4cm}\\
	\caption{\footnotesize 
	Evolution of $\gamma$-value in generations for 
	the topological model.}
	\label{fig:gamma_evolution_topo}
\end{figure}
\mbox{}

On the other hand, 
for the topological model, 
we find the order 
of interactions as $J_{2} > J_{3} > J_{4} > J_{1}$ as 
shown in Fig. \ref{fig:gamma_evolution_topo}. 
\subsection{Theoretical upper bound of the $\gamma$-value}
In our simulations, we used the $\gamma$ value as a `cost' for 
the optimization problem to determine the three interactions in BOIDS. 
This procedure is possible because 
the $\gamma$-value itself is described as a function 
of these interactions implicitly, namely, 
$\gamma = \gamma (J_{1},J_{2},J_{3})$. 
Thus, by using the GA (the choice of GA as a tool to maximize it is 
not essential here, and of course,  one can use the different 
methods such as simulated annealing), 
we maximized it under the `zero-collisions' and 
`no breaking-up' constraints. 
In this sense, this simple maximization 
process under two constraints leads to 
the empirical value of $\gamma$. 
Therefore, we might conclude that 
the flock is formed so as to maximize the $\gamma$-value 
and to satisfy the two constraints. 

In fact, it is obvious from the definition that the upper bound of 
the $\gamma$-value is $1$. 
However, from the result shown in Table.\ref{tbl:GA_result}, 
the $\gamma$-value observed in our simulations is lower than the 
bound, namely, $\gamma \simeq 0.8$.   
We also examined to what extent the $\gamma$-value increases 
when we does not require any `zero-collision' constraint during 
the GA dynamics and 
found that the $\gamma$ value increases up to near the bound. 
From these results we may conclude that 
the `zero-collision' constraint reduces the $\gamma$-value considerably. 
However, we should notice that 
$\gamma$-value calculated 
in the empirical findings 
by Ballerini {\it et. al.} \cite{Ballerini} also 
takes the value around $0.8$  which is very close to our result. 
 Therefore, we might conclude that 
 finding the optimal weights for the interaction 
 by maximizing 
 the $\gamma$-value under `zero-collision' constraint 
 is a reasonable way to realize more `realistic' flocking simulations 
 even in our personal computers. 
 
Mathematically, 
inspired by 
the so-called {\it Gardner's capacity} in 
the research field of neural networks \cite{Gardner}, 
we might examine the following 
 {\it fraction of solution space}: 
 \begin{eqnarray}
 \nu (\gamma)  =  
 \frac{
 \int_{0}^{\infty}
 d\mbox{\boldmath $J$}
 \Theta (|\mbox{\boldmath $J$}|-j)
 \delta (N_{\rm c}) \delta (N_{\rm b}) \delta(\gamma(\mbox{\boldmath $J$})-\gamma)}
 {
 \int_{0}^{\infty}
 d\mbox{\boldmath $J$}
 \Theta (|\mbox{\boldmath $J$}|-j)
 }
 \end{eqnarray}
 where $\Theta(\cdots)$ and 
 $\delta (\cdots)$ stand for 
 a step function and a delta function, respectively. 
 Here we also defined  
 $d\mbox{\boldmath $J$} \equiv 
 dJ_{1}dJ_{2}dJ_{3}$, 
 $\Theta (|\mbox{\boldmath $J$}|-j) \equiv 
  \Theta(|J_{1}|-j_{1}) \Theta(|J_{2}|-j_{2})  \Theta(|J_{3}|-j_{3})$. 
$N_{c}$ and  $N_{b}$ mean the numbers of collisions and breaking-up. 
Fixed constant variables 
$j_{k}, k=1,2,3$ specify the supports for these 
three variables $J_{k}, k=1,2,3$.  
The fraction  $\nu$ might shrink (perhaps, monotonically) to zero when 
we increases the $\gamma$ and 
a `non-trivial' theoretical upper bound $\gamma_{c}$ 
might be obtained as a solution of $\nu (\gamma_{c})=0$. 
In our forthcoming article, the details of this argument 
and the results will be reported. 
\subsection{Symmetry breaking in the space of interactions}
In our BOIDS modelling, 
we gave the interactions 
such as $J_{1},J_{2}$ and $J_{3}$ 
for the metric model ($J_{1},J_{2},J_{3}$ and $J_{4}$ 
for the topological model) 
to all agents as the same values. 
However, of course, 
one could modify the 
situation so as to give the 
`agent-dependent interactions' as $(J_{1}^{(i)},J_{2}^{(i)},J_{3}^{(i)})$ 
to the system. 
Then, it is very interesting for us to 
make two-dimensional histograms for each of the interactions 
like $J_{1}(\theta,\phi), J_{2}(\theta,\phi),J_{3}(\theta,\phi)$. 
From the histograms, we might obtain 
the useful information about the 
correlation (relationship) between anisotropy in position and 
anisotropy in the interaction. 
However, apparently 
the number of 
parameters to be determined by GA 
increases from three to $3N$ and 
the number (system size) is critical for our 
computational cost of 
determination by GA within a reliable precision and 
realistic computational time. 
Therefore, we would like to address this problem in 
our future studies. 
\subsection{Comparison with empirical findings}
From Fig.\ref{fig:N_Gamma_Result} and 
the same plot given by Ballerini {\it et. al.} \cite{Ballerini}, 
$\gamma$-values for $n=1$ are almost the same and 
the both decrease as $n$ increases. 
Moreover, these two curves 
(ours and Ballerini's) converges to $1/3$ in the asymptotic 
limit of  $n \to \infty$. 
However, 
in our evaluation, the $\gamma$-value becomes lower than $1/3$ 
in the range of $7 \leq n \leq 14$, whereas the empirical evidence indicates 
$\gamma$-value monotonically decreases as $n$ increases and 
is saturated to $1/3$ beyond $n=6$. 
This is an essential difference between the Ballerini's empirical findings and ours. 

Here, we assume that the metric interaction might cause a 
regular-polygon structure (a locally crystallized structure)
 in the flocking leading up to the `anti-anisotropy'. 
Hence, in this paper, we also 
carried out the BOIDS simulations 
in which the neighbours for 
an arbitrary agent is defined topologically instead of metrically. 
We found that the anisotropy measurement 
does not show the anti-anisotropy and 
the gap between our modelling and empirical findings was actually reduced. 
However, 
the $\gamma$-value for $n=1$ 
is apparently smaller than the 
empirical evidence (see Fig.\ref{fig:result_top} (left)) and 
it might be needed to introduce different 
types of constraints to design the 
artificial flocking in computers. 
\section{Summary}
In this paper, we utilized GAs to maximize the $\gamma$-value 
in order to determine the weights for interactions in the BOIDS under 
`zero-collision' and `no-breaking-up' constraints. 
We found that this procedure enables us to simulate the 
realistic flocking phenomena even in our personal computer level. 
We showed that  the resultant $\gamma$-value 
as a function of $n$-th order of the nearest neighbouring agents is 
quite similar to the empirical findings in several aspects \cite{Ballerini,Cavagna}. 
We carried out the simulations 
for both metric and topological definitions of 
neighbours in the flocking and 
found that the topological 
model reproduce the 
physical quantities of 
empirical evidence  
much better than the 
metric model does. 
Of course, there still exists a gap between the empirical evidence and 
our results by the BOIDS modelling. 
Recently,  Bode {\it et. al.} \cite{Wood} measured this anisotropy in artificial flockings  
by using stochastic, asynchronous updating scheme for each agent's movement.
Their model partially succeeded in reproducing the behaviour of the empirical findings 
and showed us a possibility that there exists more realistic flock
simulations than the conventional deterministic one. 
In our modelling, such probabilistic ingredients were not taken into account, 
however, their numerical evidence might stress that 
such a stochastic agents is one of essential factors to generate 
much more realistic collective behaviour of flockings. 
The extensive studies to reduce the gap should be needed to 
reveal the nature of non-trivial collective behaviour  in flocking phenomena. 
%
\section*{Acknowledgement}
We were financially supported by 
Grant-in-Aid for Scientific Research (C) 
of Japan Society for 
the Promotion of Science, No. 22500195. 
The present authors thank  
anonymous referees for 
critical reading of the manuscript and useful comments. 
\section*{Appendix}
\appendix
\section{Details of our genetic algorithm}
\label{app:A}
We here explain our GA procedure in this paper as follows. 
\begin{enumerate}\setlength{\topsep}{2pt}\setlength{\partopsep}{0pt}\setlength{\itemsep}{0pt}\setlength{\parsep}{0pt}
	\def\theenumi{\roman{enumi}}	
	\item[1.] {\bf Initialization}: Create gene configurations $U$.
	\item[2.]  Repeat the following procedure ($g=1:g \le G$).
		\begin{enumerate}
		\setlength{\topsep}{0pt}
 		\setlength{\partopsep}{0pt}
		\setlength{\itemsep}{0pt}
 		\setlength{\parsep}{0pt}
		\def\theenumii{\arabic{enumii}}
			\item{\bf  Crossover}: To make a crossover to generate a new gene configuration $E$.
			\item{\bf Mutation}: Define a gene configuration $I=U+E$ and 
			pick up $|I_{\rm s}|+|I_{\rm a}|$ components randomly from $I$ and mutate them.
				
			\item{\bf Selection}: Select the gene configurations having high fitness values from $I$. 
				We select such  
				$|U|$ gene configurations and update the previous set  $U$. 
		\end{enumerate}
\end{enumerate}
We next show the above {\bf Initialization} more precisely as follows. 
\mbox{}\\
\\
\noindent{\bf Initialization:}
\begin{enumerate}\setlength{\topsep}{0pt}\setlength{\partopsep}{0pt}\setlength{\itemsep}{0pt}\setlength{\parsep}{0pt}
	\def\theenumi{\roman{enumi}}
	\item[(i1)] We give a random variable in the range $[0.001,0.999]$ to each gene $J_{1}, 
	J_{2}$ and $J_{3}$ and 
	generate a gene configuration. 
	\item[(i2)] For the generated gene configuration, we evaluate the $\gamma$-value. 
	\item[(i3)] If the $\gamma$-value is no more than $\gamma_{\rm max}^{(0)}$ and 
	any `collision' or `breaking-up' is not observed,
	we choose the $\gamma$-value as the fitness and add the corresponding 
	gene configuration $(J_{1},J_{2},J_{3})$ to the set $U$. 
	\item[(i4)] Repeat the above (i1),(i2) and (i3) until the number of configurations reaches $|U|$.
\end{enumerate}
Where we defined the $\gamma_{\rm max}^{(0)}$ to confirm that 
one can get optimal weights even if
he (or her) starts  the GA from wrong initial gene configurations having 
relatively low fitness values.
We should keep in mind that 
the evaluation of $\gamma$-value is carried out for 
$8$-independent runs of simulations and 
we set the $\gamma$-value which is averaged over $8$-independent runs 
to the fitness function if and only if there is no `collision' or `breaking-up' during each trial. 

We next explain the details of the {\bf Crossover}. \\
\\
\noindent{\bf Crossover:}
\begin{enumerate}\setlength{\topsep}{0pt}\setlength{\partopsep}{0pt}\setlength{\itemsep}{0pt}\setlength{\parsep}{0pt}
	\def\theenumi{\roman{enumi}}
	\item[(c1)] Pick up arbitrary two gene configurations from the 
	set $U$ and define these configurations as $a$ and $b$. 
	\item[(c2)] We swap arbitrary one gene in the $a$ for arbitrary one gene in the $b$. 
	\item[(c3)] We evaluate the $\gamma$-value for the modified  $a,b$ and add 
	them to the set  $E$ if and only if there is no `collision' or `breaking-up'. 
	\item[(c4)] Repeat the above procedures (c1),(c2) and (c3) until we have $|E|$ gene configurations. 
\end{enumerate}
As the searching (solution) space is constructed by 
only three variables $J_{1},J_{2}$ and 
$J_{3}$ having real numbers, 
it is naturally assumed that 
the effect of the crossover is relatively weak 
on the optimization by GAs. 
Thus, we overcome this weakness by 
introducing the following 
two kinds of {\bf Mutations}. \\
\\
\noindent{\bf Mutation}
\begin{enumerate}\setlength{\topsep}{0pt}\setlength{\partopsep}{0pt}\setlength{\itemsep}{0pt}\setlength{\parsep}{0pt}
	\def\theenumi{\roman{enumi}}
	\item[(m1)] From the set $I \equiv U+E$, we pick up a gene configuration and define it as $c_{s}$. 
	\item[(m2)] We update arbitrary one gene, say, $J_{i}$,  in the configuration $c_{s}$ randomly as 
	           $J_{i} \to J_{i} +\delta$, where $\delta$ stands for a random number in the range $[0.001, 0.01]$. 
	\item[(m3)] For the modified $c_{s}$, we evaluate the $\gamma$-value and add the $c_{s}$ to 
	the set $I$ if and only if there is no `collision' or `breaking-up'. 
	\item[(m4)] Repeat the above (m1),(m2) and (m3) until we obtain $|I_{\rm s}|$ gene configurations. 
	\item[(m5)] We pick up an arbitrary gene configuration and replace a single gene $J_{i}$ 
	with a random number in the range $[0.001, 0.999]$. 
	We make $|I_{\rm a}|$ gene configurations by making use of the same procedure and add 
	the set $I_{\rm a}$ to the set $I$. 
\end{enumerate}
It should be noted that 
the above procedures (m1),(m2) and (m3) achieves 
`local search' whereas 
the procedure (m5) acts as `global search'. 
Thus, the above {\bf Mutation} realizes 
the effective searching by using a mixture of 
local and global searches.  

Finally, we explain the details of the {\bf Selection} as follows. \\
\\
\noindent{\bf Selection}
\begin{enumerate}\setlength{\topsep}{0pt}\setlength{\partopsep}{0pt}\setlength{\itemsep}{0pt}\setlength{\parsep}{0pt}
	\def\theenumi{\roman{enumi}}
	\item[(s1)] Pick up a gene configuration having the highest $\gamma$-value from the set $I$ as an 
	{\it elite} and the evaluate the $\gamma$-value if and only if there is no `collision' or `breaking-up'. 
	Then, we add the gene configuration to the set $U$ in the next generation. 
	\item[(s2)]  
	For all genes in the set $I$, we calculate the $\gamma$-values as the corresponding fitness values.  
	If  $g>G/2$, we make a linear transform on the $\gamma$-value for all gene configurations 
	in the set $I$ and choose the transformed set of $\gamma$-values as fitness functions. 
	\item[(s3)] Make a roulette selection of the gene configuration based on the fitness functions and 
	evaluate the $\gamma$-value if and only if there is no `collision' or `breaking-up'. 
	Then, we add the gene configuration to the set $U$ and repeat the above procedure 
	until we have $|U|$ configurations. 
\end{enumerate}
As we are restricted ourselves to 
the case in which the number of trials, 
the time of observation are limited, 
there exists a possibility that we get unexpected high $\gamma$-value 
and the gene configuration having such high $\gamma$-value accidentally 
and to make matter worse, 
such gene configuration sometimes might survive until the final generation.  
To overcome this difficulty, 
we repeat the measurement of the $\gamma$-value for the 
selected gene configurations until `zero-collision' condition 
is strictly satisfied and we 
delete the selected gene configurations 
if they lead to `collision' or `breaking-up' 
even if they show the high $\gamma$-value. 
Thus, we systematically solve the optimization 
problem under two essential constraints. 
\subsection{The size of GA in computer simulations}
We next explain the size of GA in 
our computer simulations. 
We summarize them in Table \ref{tbl:GA}.  
\begin{table}[t!]
\begin{center}
\begin{tabular}{l|c}
	\hline
	\hline
	Number of generation ($G$) & 25\\
	Size of gene configurations ($|U|$) & 10\\
	Initial $\gamma$-max & 0.5\\
	Crossover rate ($|E|$) &  1\\
	Mutation rate 1 ($|I_{\rm s}|$) & 2\\
	Mutation rate 2 ($|I_{\rm a}|$) & 1\\
	\hline
	\hline
\end{tabular}
\end{center}
\caption{\footnotesize 
Parameter setting for our GA.}
\label{tbl:GA}
\end{table}
We carry out the GA having the above setting-up 
of the parameters. 
Then, for each generation of 
the GA, we evaluate the $\gamma$-value, 
the weight of the interactions and the distribution 
of the gene configurations 
for three independent  runs. 
\section{Border bias effect and procedure to avoid it}
\label{app:B}
In computer simulation for finite size systems ($N \ll \infty$), 
we should keep in mind that 
the results are sometimes influenced by 
the so-called {\it border bias effect} \cite{Cavagna2}. 
The border bias effect comes from the asymmetric shape of 
flockings. For instance, 
the asymmetric shape we mentioned here is typically 
an elliptical shape which is obtained as a deviation from 
a given symmetric sphere. 
The major axis of the elliptic corresponds to the 
direction of moving of the flock. 
For a given flocking and a given agent, 
the $n$-th nearest neighbouring agent  
is more likely to exist in the direction of 
moving of the flocking rather than the direction perpendicular to 
the flock's movement. 
As the result, 
the $\gamma$-value decreases to below $1/3$ as $n$ increases, 
 and then,  `anti-anisotropy' emerges. 
This phenomenon is nothing but border bias effect we mentioned here. 
It is very important for us 
to avoid the border bias effect to evaluate the $\gamma$-value precisely. 
If there is no correlation between the direction of flock's moving and 
the shape of the flocking, 
one may cancel the effect by taking the average of 
the $\gamma$-value over several (usually, a lot of ) 
trials with different initial conditions. 
However, if not, it is very difficult for us to cancel the effect by 
this simple procedure. 
In most cases of simulations, 
the shape of flockings is an elliptic in which 
the major axis is always in the direction of flock's moving 
and there exists an apparent correlation between the shape and the direction of moving.
In general, the effect is more serious for the flocking simulation with 
small number of agents than that with huge number of agents. 
Ballerini and Cavagna {\it et. al.} cancelled the effect by 
excluding the agents on the border, however, 
the usefulness of their procedure is limited to 
the case in which the number of mates is larger than $400$ \cite{Ballerini,Cavagna2}. 

From this fact in mind, 
we provide two distinct procedures 
to cancel the two types of border bias, namely, 
{\it multiple trial method} for border bias caused by `cubic shape' (slab) of the flocking and 
{\it sphere extraction method} for border bias induced by  `spherical shape' of the flock.   
we show that these procedures 
make our flock simulations free from 
the border bias effects. 
\subsection{Multiple trial method}
\label{subsec:B1}
We first examine the extreme situation which 
shows `fake'-anisotropy by border bias \cite{Cavagna2}. 

\begin{figure}[ht]
\begin{minipage}{0.49\linewidth}
	 \includegraphics[width=\linewidth,bb= 61 255 572 580]{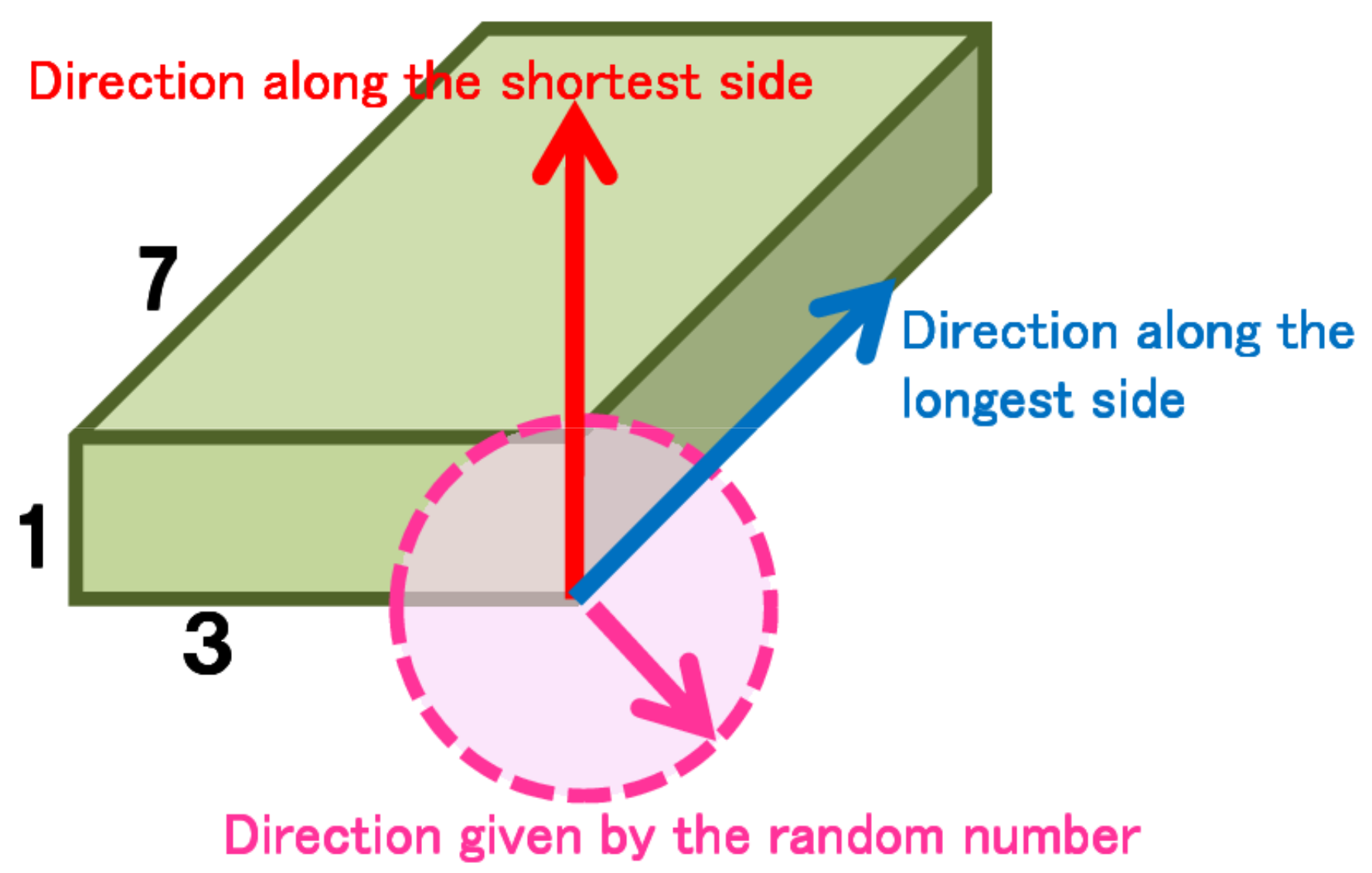}
\end{minipage}
\begin{minipage}{0.49\linewidth}
	 \includegraphics[width=\linewidth,bb=77 251 521 579]{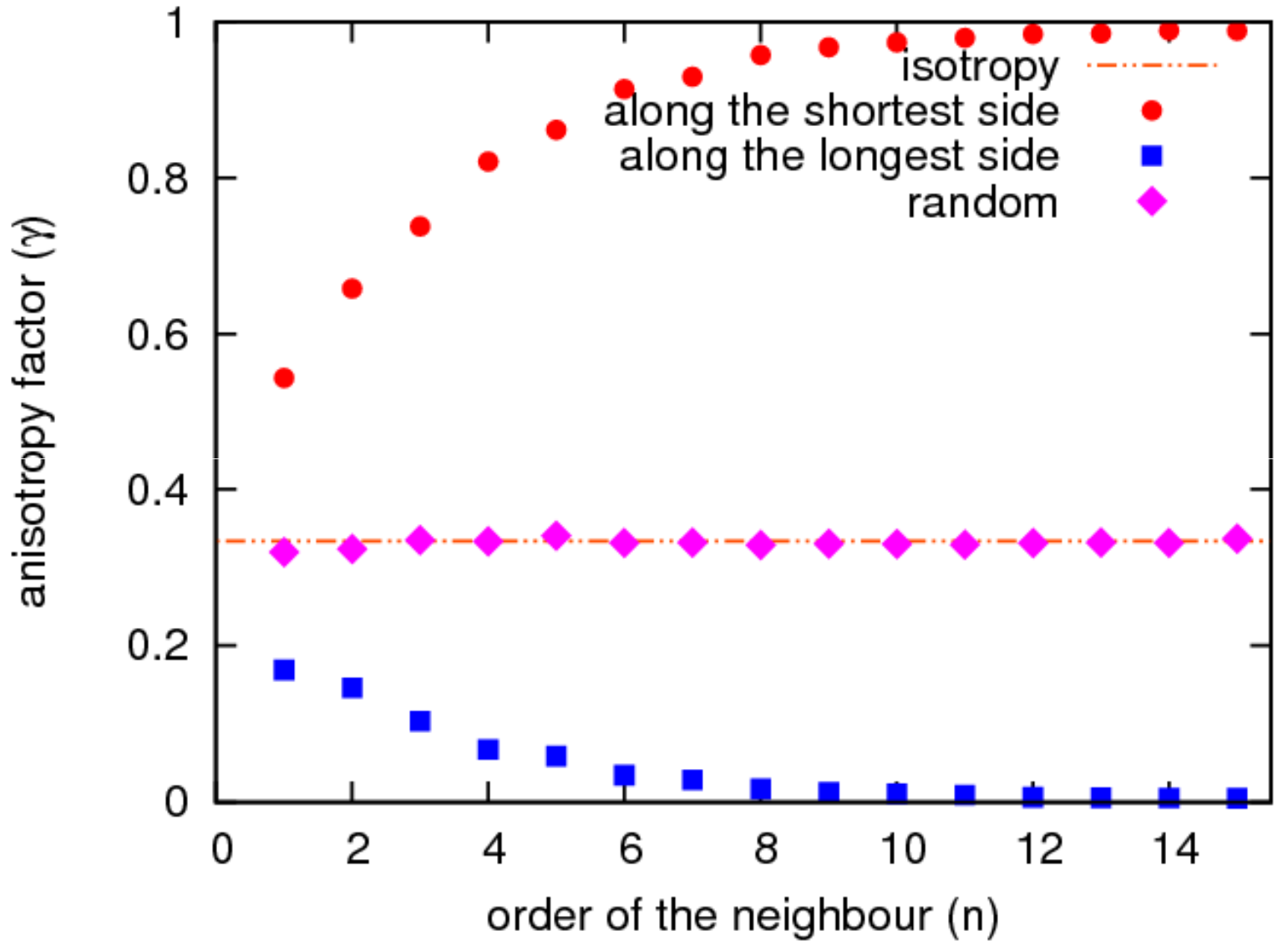}
\end{minipage} \vspace{4cm} \\
	\caption{\footnotesize 
	The extreme shape (`cubic shape') of artificial flocking 
	which shows `fake'-anisotropy (left). 
	The right panel shows 
        the $\gamma$-value as a function of order $n$. 
        Circles and boxes correspond to 
        the $\gamma$-values observed in the flocking moving to 
        the directions of the shortest and the longest sides, respectively. 
        We purposely put $1200$-individuals randomly into this cubic 
        so as to show the isotropy $\gamma=1/3$ by hand, however,
        the anisotropy/`anti-anisotropy' emerges. 
        The `diamonds' in this panel show the result by 
        the multiple trial method. We clearly find that 
        the isotropy is observed as we expected. 
        }
	\label{fig:border1}
\end{figure}

Let us think about the artificial flocking having a cubic shape 
as shown in the left panel of Fig.\ref{fig:border1}. 
The ratio of three slides of the cube is given as $7:3:1$. 
Then, we evaluate the $\gamma$-value 
for the flock moving to the direction of the shortest side.  
We also purposely make the flocking so as to 
show the isotropy for the angular distribution 
for individual by hand. 
Hence, we should naturally observe the `isotropy' through 
the $\gamma$-value, namely, 
$\gamma=1/3$ if one correctly simulates the artificial flocking 
without any border bias effect.  

The result is shown in the right panel of Fig. \ref{fig:border1} by `circles'. 
From this panel, we clearly find that the 
anisotropy emerges and our simulations are 
apparently affected by the border bias. 
Even if we change the direction of flock's motion 
from the shortest side to the longest slide, 
one cannot obtain the `isotropy' but one observes  
the `anti-anisotropy' as shown  in the right panel of 
Fig. \ref{fig:border1} by `boxes'. 
\mbox{}

In order to overcome this type of border bias effects, 
we utilize the so-called {\it multiple trial method}. 
Namely, we evaluate the measurements such as 
the $\gamma$-value for a lot of independent trials, in each of 
which we give the initial direction of 
flock's motion randomly without any correlations 
with any specific direction such as 
the shortest and the longest directions of cubic.
Then, the measurement is obtained 
as an average over the trials. 
We show the resulting $\gamma$-value for $1000$-trials in 
the right panel of Fig. \ref{fig:border1} 
by `diamonds'. 
We clearly find that 
the `isotropy' is observed as we expected.  
\subsection{Sphere extraction method}
\label{subsec:B2}
In the previous subsection \ref{subsec:B1}, we were confirmed that 
the multiple trial method efficiently 
reduces the border bias effect. 
However, 
we should notice that 
the method is effective only for the case 
in which the direction 
of flock's movement (specified by the velocity 
vector $\mbox{\boldmath $V$}$) 
and the shape of the flocking has no correlation. 

To understand it, let us 
define a unit vector pointing 
to the direction of 
the longest side of the cubic in the previous subsection 
\ref{subsec:B1}  
by $\mbox{\boldmath $e$}_{L}$. 
Then, the following 
condition should be satisfied 
\begin{eqnarray}
\overline{
\mbox{\boldmath $e$}_{L} \cdot 
\mbox{\boldmath $V$}} & = & 0 
\end{eqnarray}
to use the multiple trial method effectively, 
where $\overline{(\cdots)}$ 
stands for the time average over 
the sampling from BOIDS dynamics. 
It is obviously confirmed when 
we consider the extreme case in which 
randomly given vectors $\mbox{\boldmath $V$}_{k}$ ($k=1,\cdots, M$: $M$ is the number 
of trials) as 
the direction of flock's motion are
always correlated with 
$\mbox{\boldmath $e$}_{L}$ 
in terms of $\overline{\mbox{\boldmath $e$}_{L} \cdot 
\mbox{\boldmath $V$}_{k}}  =1\, (k=1,\cdots, M)$. 
For this case, we clearly notice that 
the multiple trial method gives 
exactly the same result as 
the `circles' shown in the right panel of Fig. \ref{fig:border1}. 

Therefore, another way to 
reduce the border bias effect is needed to 
evaluate the anisotropy measurement 
precisely in our BOIDS simulations.

\begin{figure}[!t]
	\centering
	\includegraphics[width=0.9\linewidth,bb=0 0 509 366]{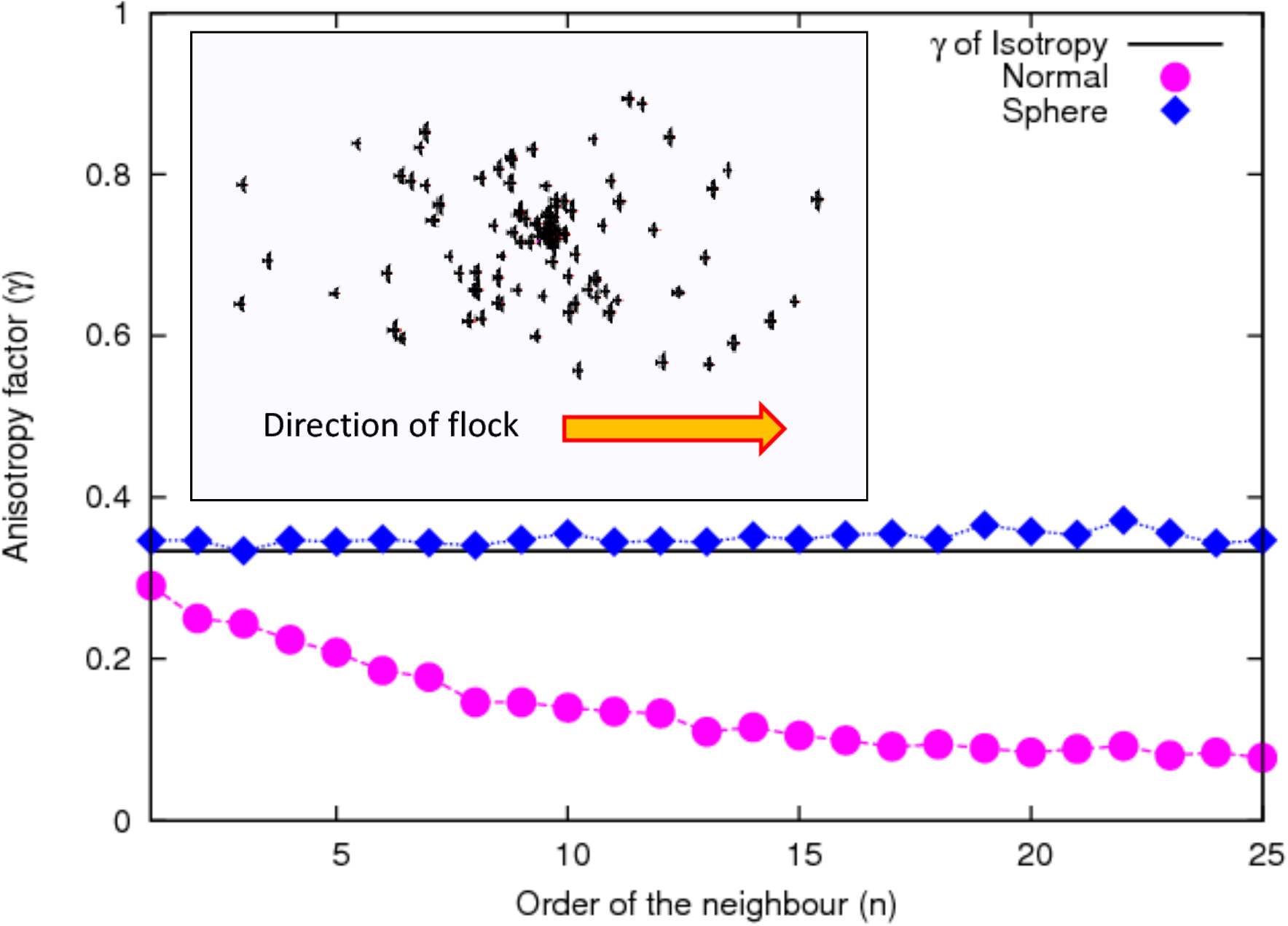}	
	\caption{\footnotesize 
	The $\gamma$-value as a function of order $n$.  A typical snapshot of the aggregation 
	is shown in the inset.}
	\label{fig:border}
\end{figure}

In order to reduce the border bias effect in such cases,
we take only agents who are in the sphere being 
inscribed with the shape of 
the flocking to calculate the $\gamma$-value. 
To check the usefulness of our procedure, 
we set-up $100$ agents moving with 
the same velocity in the same direction and  
give them their initial positions uniformly in the elliptic 
in which the major axis is in the direction of 
flock's moving (see the inset of Fig. \ref{fig:border}). 
For these agents, we evaluate the 
$\gamma$-value up to $n=25$ and 
compare the result with 
the $\gamma$-value calculated without the above procedure. 
The results are shown in Fig.\ref{fig:border}. 
We carried out $1000$-independent runs and 
calculated the average of the $\gamma$-value. 
As we mentioned, the distribution of 
the position of agents is uniform and 
the correct (true) $\gamma$-value should be $1/3$. 
From this figure, we find that our procedure 
(circle {\sf Sphere})  
works well in comparison with 
the result without the procedure (diamond {\sf Normal}).



\begin{thebibliography}{99}


\bibitem{modelling}
Iwo Bialynicki-Birula and 
Iwona Bialynicka-Birula, 
{\it Modeling Reality: 
How computers mirror life}, 
{\it Oxford University Press} 
(2004).


\bibitem{Inada}
Y. Inada, K. Kawachi, 
{\it Order and flexibility in the motion of fish schools}, 
{\it J. of Theoretical Biology}
 {\bf 214}, pp.371-387 (2002).


\bibitem{Okubo}
A. Okubo
{\it Dynamical aspects of animal 
grouping: Swarms, schools, flocks, and herds},
{\it Advances in Biophysics} {\bf 22},
pp.1-94 (1986).


\bibitem{Landau}
D.P. Landau and K. Binder, 
{\it A Guide to Monte 
Carlo Simulations in 
Statistical Physics}, 
{\it Cambridge University Press} (2005).



\bibitem{Vicsek}
T. Vicsek, A. Czir$\acute{\rm o}$k, 
E. Ben-Jacob, I. Cohen and O. Shochet, 
{\it Novel type of phase
transition in a system of self-driven particles},
{\it Physical Review Letters} {\bf 75},
pp.1226-1229 (1995). 


\bibitem{Reynolds}
C.W. Reynolds, 
{\it Flocks, Herds, and Schools: A Distributed Behavioral Model}, 
{\it Computer Graphics} {\bf 21}, 
pp.25-34 (1987). 

\bibitem{BOIDS}
{\tt http://www.red3d.com/cwr/boids/}



\bibitem{Tanner}
H. G. Tanner, A. Jadbabaie, and G. J. Pappas, {\it Stable flocking of mobile
agents, part I: Fixed topology}, 
{\it Proceedings of  the 42nd IEEE Conference of 
DecisionControl}, pp.2010- 2015 (2003).


\bibitem{Tanner2}
H. G. Tanner, A. Jadbabaie, and G. J. Pappas, {\it 
Stable flocking of mobile agents, part II: Dynamic topology}, 
{\it Proceedings of  the 42nd IEEE Conference of 
DecisionControl}, pp. 2016-2021 (2003).



\bibitem{Olfati}
R.  Olfati-Saber, 
{\it Flocking for Multi-Agent Dynamic Systems:
Algorithms and Theory}, 
{\it IEEE Transactions on Automatic Control}
 {\bf 51}, No. 3, pp.401-420 (2006). 



\bibitem{Aoki}
I.  Aoki, 
{\it A simulation study on the schooling mechanism in fish}, 
{\it Bulletin of the Japanese Society of Scientific Fisheries}
{\bf 48}, pp.1081-1088 (1982). 


\bibitem{Su}
H. Su, X. Wang, Z. Lin, 
{\it Flocking of multi-agents with a virtual leader}, 
{\it  IEEE Transactions on 
Automatic Control} {\bf 54},
No. 2, pp.293-307 (2009).


\bibitem{Herbert}
H. G. Tanner, A. Jadbabaie and G. J. Pappas, 
{\it Flocking in Fixed and Switching Networks}, 
{\it  IEEE Transactions on Automatic Control}
{\bf 52}, No 5, pp.863-868 (2007).



\bibitem{Ballerini}
M. Ballerini, N. Cabibbo, R. Candelier, A. Cavagna, 
E. Cisbani, I. Giardina, V. Lecomte, A. Orlandi, 
G. Parisi, A. Procaccini, M. Viale and V. Zdravkovic, 
{\it Interaction Ruling Animal Collective 
Behaviour Depends on Topological rather than Metric Distance, 
Evidence from a Field Study}, 
{\it Proceedings of the National Academy of Sciences USA}
 {\bf 105}, No. 4, pp.1232-1237 (2008). 

\bibitem{Ballerini2}
M. Ballerini, N. Cabibbo, R. Candelier, A. Cavagna,
E. Cisbani, I. Giardina,  A. Orlandi,
G. Parisi, A. Procaccini, M. Viale and V. Zdravkovic,
{\it An empirical study of large, naturally occurring starling flocks:
 a benchmark in collective animal behaviour} 
{\it Animal Behaviour}
{\bf 76}, pp.201-215 (2008).

\bibitem{Makiguchi}
M. Makiguchi and J. Inoue, 
{\it Numerical Study on the Emergence of Anisotropy in Artificial Flocks: 
A BOIDS Modelling and Simulations of Empirical Findings},
{\it Proceedings of the Operational Research Society 
Simulation Workshop 2010 (SW10), CD-ROM}, pp. 96-102
 (the preprint version, arxiv:1004 3837) (2010). 


\bibitem{Dirac}
P.A.M. Dirac, 
{\it The principles of quantum mechanics }, 
{\it Clarendon Press, Oxford} (1930). 



 
\bibitem{Goldberg}
D. E. Goldberg, 
{\it Genetic Algorithms in Search, Optimization and 
 Machine Learning}, 
{\it Addison Wesley} (1989). 

\bibitem{Chen}
Y-W. Chen, 
K. Kobayashi, 
H. Kawabayashi and X. Huang, 
{\it Application of Interactive Genetic Algorithms to Boid Model Based Artificial Fish Schools}.
{\it Lecture Notes in Artificial Intelligence, Springer}
{\bf 5178}, pp.141-148 (2008). 


\bibitem{Gardner}
E. Gardner, 
{\it The space of interactions in neural network models},
{\it Journal of Physics A: Mathematical and General}
{\bf 21}, pp.257-270 (1988).

\bibitem{Cavagna}
A. Cavagna, 
A. Cimarelli, I. Giardina, 
G. Parisi, R. Santagati, F. Stefanini
and M. Viale, 
{\it Scale-free correlations in starling flocks}, 
{\it Proceedings of the National Academy of Sciences USA} 
{\bf 107}, No. 26,  pp. 11865-11870 (2010). 

\bibitem{Cavagna2}
A. Cavagna,
I. Giardina, A. Orlandi, G. Parisi, A. Procaccini,
M. Viale and V. Zdravkovic,
{\it The STARFLAG handbook on 
collective animal behaviour: Part II},
{\it Animal Behaviour}
{\bf 76},
Issue 1, pp.237-248 (2008).


\bibitem{Wood}
N.W.F. Bode, D.W. Franks and A. Jamie Wood, 
{\it Limited interactions are necessary for realistic movement in animal groups}, 
{\it Proceedings of the Royal Society Interfaces}, 
published online before print September 8, 2010, doi:10.1098/rsif.2010.0397 (2010). 

\bibitem{Cavagna2008}
A. Cavagna, 
A. Cimarelli, 
I. Giardina,  
A. Orlandi, 
G. Parisi, 
A. Procaccini, 
R. Santagati and F. Stefanini, 
{\it New statistical tools for analyzing the structure of animal groups},
{\it Mathematical Biosciences} 
{\bf 214}, pp.32-37 (2008). 




\end{thebibliography}
\end{document}